\documentclass[10pt,a4paper]{article}
\usepackage[utf8]{inputenc}
\usepackage{amsmath}
\usepackage{amsfonts}
\usepackage{amssymb}
\usepackage{graphicx}

\newtheorem{theorem}{Theorem}

\newtheorem{definition}[theorem]{Definition}
\newtheorem{proposition}[theorem]{Proposition}

\newtheorem{example}[theorem]{Example}

\usepackage{hyperref}

\usepackage[left=3cm,right=3cm,top=3cm,bottom=3cm]{geometry}

\title{\textbf{A new  robust approach for the polytomous logistic regression model based on Rényi's pseudodistances}}
\author{Elena Castilla \footnote{Departamento de Matematica Aplicada, Rey Juan Carlos University, Mostoles Campus, 28933 Madrid, Spain. elena.castilla@urjc.es}}
\date{}
\begin{document}
\maketitle
\begin{abstract}
This paper presents a robust alternative to the Maximum Likelihood Estimator (MLE) for the Polytomous Logistic Regression Model (PLRM), known as the family of minimum Rènyi Pseudodistance (RP) estimators.  The proposed minimum RP estimators are parametrized by a tuning parameter $\alpha\geq0$, and include the MLE as a special case when $\alpha=0$. These estimators, along with a family of RP-based Wald-type tests, are shown to exhibit superior performance in the presence of misclassification errors. The paper includes an extensive simulation study and a real data example to illustrate the robustness of these proposed statistics. 
\end{abstract}


\section{Introduction}

The polytomous logistic regression model (PLRM), also known as multinomial logistic regression model, is as a natural extension of the classical logistic regression when the response variable has more than two categories. This technique has been successfully used for both classification and inference involving an unordered categorical response and explanatory variables across various fields. For instance, it has been utilized in geographical studies for soil mapping (Piccini et al., 2019), in behavioral education to forecast adolescent risk (Peng and Nichols, 2003), in social service research (Petrucci, 2009), in investigations of physical activities (Gander, 2011; Novas et al., 2003), and in medical research (Manor et al., 2000; Biesheuvel, 2008), among many others.
 
Most of the results mentioned above are based on the Maximum Likelihood Estimator (MLE), which is well-known for its efficiency but also for its lack of robustness. To address this issue, some robust approaches have been proposed in the literature. Castilla et al. (2018) derived estimators and test statistics based on the Density Power Divergence (DPD) to handle misclassification errors, while Miron et al. (2022) introduced two new estimators to cope with outlying covariates. In this article, we present an alternative robust generalization of the MLE for the  PLRM  by using the Rènyi Pseudodistance (RP) of Jones et al. (2001).  RP-based inference has gained popularity recently, especially due to its strong robustness properties against outliers, as demonstrated by Broniatowski et al. (2012) and Castilla et al. (2022a, 2022b), among others.

The paper is organized as follows. We begin in Section \ref{sec:model} by introducing the PLRM and discussing classical inference based on the MLE.  In Section \ref{sec:RP}, we define the family of minimum RP estimators, which includes the MLE as a special case, and derive their estimating equations and asymptotic distribution. In Section \ref{sec:Wald}, we present a family of Wald-type tests that generalize the MLE-based Wald test. In Section \ref{sec:IF}, we derive their influence functions. The robustness of the proposed statistics against misclassification is empirically illustrated through an extensive simulation study in Section \ref{sec:MC} and a real data example in Section \ref{sec:Data}. Finally, we conclude with some remarks in Section \ref{sec:FR}.

\section{Model formulation \label{sec:model}}
Let us consider that we have $n$ independent observations $(\boldsymbol{x}_i,\boldsymbol{y}_i)$, $i=1,\dots,n$,  where $\boldsymbol{y}_i=(y_{i1},\dots,y_{i,d+1})^T$ is a categorical outcome variable with $d+1$ categories, with $y_{ij}=1$ if the response of the $i$-th observation lies in the $j$-th category and $y_{ij}=0$ if not.  Here,  $\boldsymbol{x}_{i}=(1,x_{i1},\dots,x_{ik})^T$ is the vector of an intercept and $k$ predictors  associated to the $i$-th observation. To explain the dependence of $\boldsymbol{y}_i$ on the explanatory variables, a linear relationship is considered for each category,
\begin{align*}
\eta_{ij}=\boldsymbol{x}_i^T\boldsymbol{\beta}_j=\beta_{0j}+\beta_{1j}x_{i1}+\cdots+\beta_{kj}x_{ik}, \quad j=1,\dots,d+1,
\end{align*}
where $\boldsymbol{\beta}_j=(\beta_{0j},\dots,\beta_{kj})^T\in \mathbb{R}^{k+1}$ is the vector of unknown parameters associated to the $j$-th category. Let $\pi_j(\boldsymbol{x},\boldsymbol{\beta})$ denote the probability that the response variable belongs to the $j$-th category, $j=1,\dots,d+1$. The PLRM is a particularization of the generalized linear model (GLM) where  this probability is given by
\begin{align}\label{eq:PLRM}
\pi_{ij}(\boldsymbol{\beta})\equiv \pi_j(\boldsymbol{x}_i,\boldsymbol{\beta})=\dfrac{\exp(\eta_{ij})}{1+\sum_{s=1}^d \exp(\eta_{is})},
\end{align}
for $j=1,\dots,d$, with $\pi_{i,d+1}(\boldsymbol{\beta})=1-\sum_{j=1}^d \pi_{ij}(\boldsymbol{\beta})$. Here $\boldsymbol{\beta}=(\boldsymbol{\beta}_1^T,\dots,\boldsymbol{\beta}_d^T)^T \in \mathbb{R}^{d(k+1)}$ is the vector of unknown parameters. Let us also define the theoretical probability vector  $\boldsymbol{\pi}_i(\boldsymbol{\beta})=(\pi_{i1}(\boldsymbol{\beta}),\dots,\pi_{i,d+1}(\boldsymbol{\beta}))^T$  for each observation $i=1,\dots,n$.

\subsection{Maximum Likelihood Estimator (MLE)}

The MLE of $\boldsymbol{\beta}$, $\widehat{\boldsymbol{\beta}}_{\small \text{MLE}}$, is given by
\begin{align}
\widehat{\boldsymbol{\beta}}_{\small \text{MLE}}=\underset{\boldsymbol{\beta}}{\text{arg max}} \ \ell (\boldsymbol{\beta}),
\end{align}
where $\ell (\boldsymbol{\beta})$ is the log-likelihood of the model, given by
\begin{align}\label{eq:loglik}
\ell (\boldsymbol{\beta})=c+\sum_{i=1}^n \sum_{j=1}^{d+1} y_{ij} \log \pi_{ij}(\boldsymbol{\beta}),
\end{align}
for any positive constant $c$.  

With superscript $^{\ast}$ along with a vector (or matrix), we will denote the truncated vector (matrix) obtained by deleting the last value (row) from the initial vector (matrix). Thus $\boldsymbol{\pi}_i^{\ast}(\boldsymbol{\beta})=(\pi_{i1}(\boldsymbol{\beta}),\dots,\pi_{id}(\boldsymbol{\beta}))^T$ and $\boldsymbol{y}^{\ast}_i=(y_{i1},\dots,y_{id})^T$.  On the other hand, let $\boldsymbol{p}$ be a vector and let us denote $\boldsymbol{\Delta}(\boldsymbol{p}) = diag(\boldsymbol{p}) - \boldsymbol{p}\boldsymbol{p}^T$, where $diag(\boldsymbol{p})$ denotes the matrix with the entries of $\boldsymbol{p}$ along the diagonal.

The vector $\widehat{\boldsymbol{\beta}}_{\small \text{MLE}}$ can be obtained as the solution of the following system of equations
\begin{align}\label{eq:estimatingMLE}
\sum_{i=1}^n \boldsymbol{\Psi}_i(\boldsymbol{\beta})=\boldsymbol{0}_{d (k+1)},
\end{align}
where
\begin{align*}
\boldsymbol{\Psi}_i(\boldsymbol{\beta})=\boldsymbol{\Delta}^*(\boldsymbol{\pi}_i(\boldsymbol{\beta})) \text{diag}^{-1}\left(\boldsymbol{\pi}_i(\boldsymbol{\beta})\right)\left(\boldsymbol{y}_i-\boldsymbol{\pi}_i(\boldsymbol{\beta})\right)\otimes \boldsymbol{x}_i=\left(\boldsymbol{y}_i^{\ast}-\boldsymbol{\pi}_i^{\ast}(\boldsymbol{\beta}) \right) \otimes \boldsymbol{x}_i,
\end{align*}
where $\otimes$ denotes the kronecker product, and $\boldsymbol{0}_{d (k+1)}$ represents the null column vector of dimension $d (k+1)$.\\

 The asymptotic distribution of the MLE is given by
\begin{align}\label{eq:asymMLE}
\sqrt{n}\left(\widehat{\boldsymbol{\beta}}_{MLE}-\boldsymbol{\beta}^0\right)\overset{\mathcal{L}}{\underset{n\rightarrow \infty }{\longrightarrow }}\mathcal{N}\left( \boldsymbol{0}_{d(k+1)},\boldsymbol{\Omega}^{-1}(\boldsymbol{\beta}^0)\right),
\end{align}
where $\boldsymbol{\beta}^0$ denotes the true value of  $\boldsymbol{\beta}$ and 
$$\boldsymbol{\Omega}(\boldsymbol{\beta})=\lim_{n\rightarrow \infty}\boldsymbol{\Omega}_{n}(\boldsymbol{\beta})=\lim_{n\rightarrow \infty} \frac{1}{n}\sum_{i=1}^n\boldsymbol{\Delta}(\boldsymbol{\pi}^*_i(\boldsymbol{\beta}))\otimes \boldsymbol{x}_i \boldsymbol{x}_i^T.$$

The MLE is widely recognized for its property of being the Best Asymptotically Normal (BAN) estimator. However, it's also well-documented that this estimator lacks robustness and can exhibit significant bias due to mislabeling. To illustrate this issue, we introduce the following example.

\subsection{A real data example  \label{sec:small_example}}
\begin{figure}[h]
\center
\begin{tabular}{cc}
\includegraphics[scale=0.377]{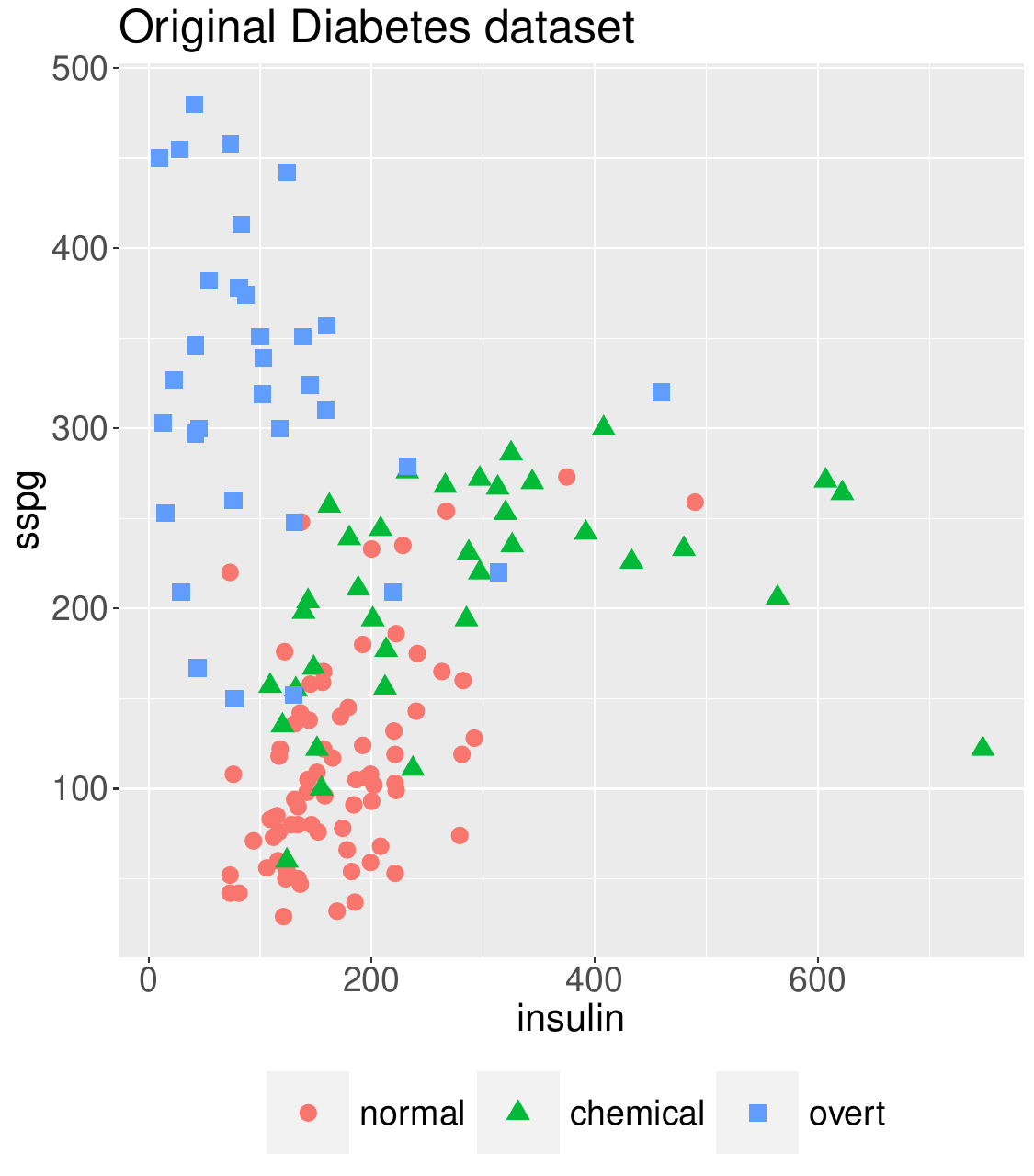}
 & \includegraphics[scale=0.377]{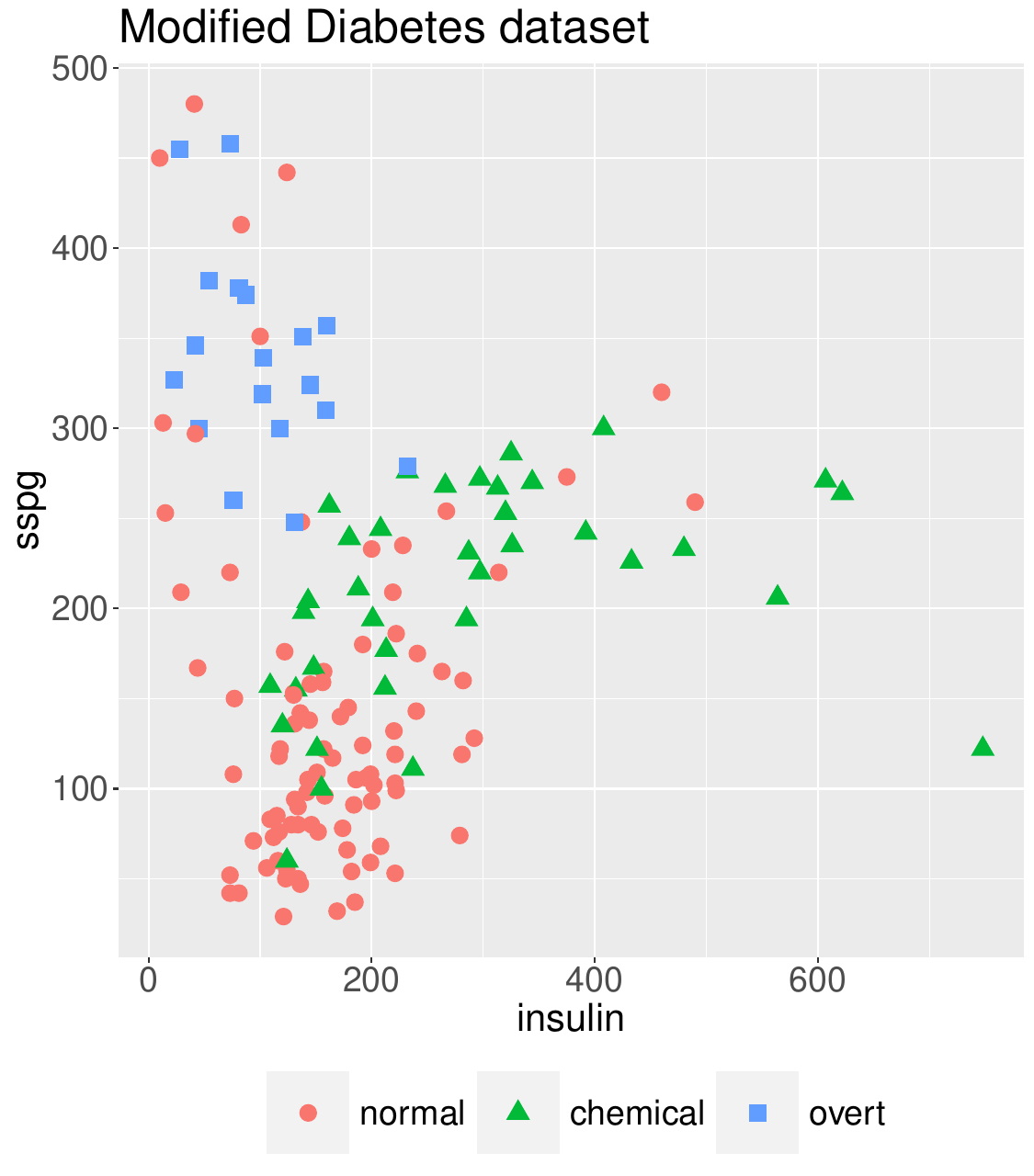}
\end{tabular} 
\caption{Diabetes Dataset: The plot shows the original dataset (left) and the modified dataset (right) obtained by modifying the categories of the last $14$ observations. \label{fig:DIABETES_example}}
\end{figure}

Let us consider the Diabetes dataset. This dataset comprises measurements from 145 non-obese adult patients, classified into three groups: \textit{normal}, \textit{chemical diabetic}, and \textit{overt diabetic}. Reaven and Miller (1979) utilized the dataset to explore the characteristics of chemical diabetes through a comprehensive, multidimensional analysis. Their work was built upon the foundational research of Friedman and Rubin (1967), who implemented a cluster analysis on the three key variables - glucose area, insulin area, and steady state plasma glucose (sspg) - leading to the identification of three distinct clusters. This dataset was also studied by Hawkins and McLachlan (1997) and recently by Iannario and Monti (2023) and is available in the R package \texttt{rrcov}. In this study, we make a specific emphasis on two measurements: sspg, which is a measure of insulin resistance, and insulin area (see Figure \ref{fig:DIABETES_example}).\\

This is a clear example of where the PLRM can be applied. We compute the MLE of model parameters and the estimated category probabilities for each of the observed subjects. To measure how the MLE adjusts these data, we assign categories based on the highest estimated probabilities and compare these to the observed ones. Only $9$ observations out of the original $145$ are wrongly classified, showing a good performance of MLE.

Now, to illustrate how misclassification errors may affect our estimation, let us assume that there has been an error in collecting our data and the $14$ last observations, which correspond to the \textit{overt} category, are wrongly collected as \textit{normal}, as illustrated in the right part of Figure \ref{fig:DIABETES_example}. In this case, if we compute the MLE and follow the previous procedure (comparing the estimated categories with the original real data), a total of $29$ observations are classified wrongly. Although this is quite natural, as we have altered the original data including some errors in the response variable, the question that arises here is if we may be able to mitigate this increment of error. In the subsequent section, we introduce the minimum RP estimator as an alternative to MLE that may lead to better estimations in this kind of situations.

\section{Minimum RP estimation  \label{sec:RP}}
Given the empirical and the theoretical probability vectors $\boldsymbol{y}_i$ and $\boldsymbol{\pi}_i(\boldsymbol{\beta})$, respectively, for each observation $i$, $i=1,\dots,n$, the RP between  $\boldsymbol{\pi}_i(\boldsymbol{\beta})$ and $\boldsymbol{y}_i$ is defined for $ \alpha>0$ as

\begin{align}
R_{\alpha}\left(\boldsymbol{\pi}_i(\boldsymbol{\beta}),\boldsymbol{y}_i \right)&=-\frac{1}{\alpha} \log  \left\{ \dfrac{\sum_{j=1}^{d+1} \pi_{ij}^{\alpha}(\boldsymbol{\beta})y_{ij}}{\left[\sum_{j=1}^{d+1} \pi_{ij}^{\alpha+1}(\boldsymbol{\beta})\right]^{\frac{\alpha}{1+\alpha}}}\right\}, \notag\\
&=\frac{1}{1+\alpha} \log \left\{\sum_{j=1}^{d+1} \pi_{ij}^{\alpha+1}(\boldsymbol{\beta}) \right\}-\frac{1}{\alpha}\log \left\{ \sum_{j=1}^{d+1} \pi_{ij}^{\alpha}(\boldsymbol{\beta})y_{ij}\right\},\label{eq:RP}
\end{align}
while for the particular case $\alpha=0$, the RP coincides with the  Kullback-Leibler divergence between  $\boldsymbol{\pi}_i(\boldsymbol{\beta})$ and $\boldsymbol{y}_i$, $d_{\text{KL}}(\boldsymbol{\pi}_i(\boldsymbol{\beta}),\boldsymbol{y}_i )$,

\begin{align}\label{eq:RP0}
R_{0}\left(\boldsymbol{\pi}_i(\boldsymbol{\beta}),\boldsymbol{y}_i \right)=d_{\text{KL}}(\boldsymbol{\pi}_i(\boldsymbol{\beta}),\boldsymbol{y}_i )=\sum_{j=1}^{d+1} y_{ij} \log \frac{y_{ij}}{•\pi_{ij}(\boldsymbol{\beta})}.
\end{align}

Note that the random variables associated with the observed response $\boldsymbol{y}_i$, given the covariate value $\boldsymbol{x}_i$ under the PLRM, are independent but non-homogeneous (let us denote this as an i.n.i.d.o. setup). Therefore, we may follow the theory by Castilla et al. (2022a),  in which the desired estimator  is obtained through the minimization of the average RP between the observed data and the model probability mass functions over each distribution. Note that the minimization of (\ref{eq:RP}) is equivalent to the maximization of 
\begin{align}\label{eq:Vi}
s_i^{(\alpha)}\left(\boldsymbol{y}_i,\boldsymbol{\beta} \right)=\dfrac{\sum_{j=1}^{d+1} \pi_{ij}^{\alpha}(\boldsymbol{\beta})y_{ij}}{\left[\sum_{j=1}^{d+1} \pi_{ij}^{\alpha+1}(\boldsymbol{\beta})\right]^{\frac{\alpha}{1+\alpha}}}.
\end{align}
On the other hand, it is not difficult to see that the minimization of the average Kullback-Leibler divergence in (\ref{eq:RP0}) is equivalent to the maximization of the log-likelihood in (\ref{eq:loglik}). We may then give the following definition.

\begin{definition}
Given the PLRM in (\ref{eq:PLRM}), the minimum RP estimator with tuning parameter $\alpha>0$ for  $\boldsymbol{\beta}$, $\widehat{\boldsymbol{\beta}}_{\alpha}$, is given by
\begin{align}\label{eq:Halpha}
\widehat{\boldsymbol{\beta}}_{\alpha}=\underset{\boldsymbol{\beta}}{\text{arg max}} \ H^{(\alpha)}_{n}(\boldsymbol{\beta}),
\end{align}
where $H^{(\alpha)}_{n}(\boldsymbol{\beta})=\frac{1}{n} \sum_{i=1}^n s_i^{(\alpha)}\left(\boldsymbol{y}_i,\boldsymbol{\beta} \right)$, and coincides with the MLE for $\alpha=0$.
\end{definition}

Castilla et al. (2022a) established the consistency of the minimum RP estimators under some standard regularity conditions for the i.n.i.d.o. setup.\\

To derive the estimating equations, we need to obtain the derivative  of (\ref{eq:Vi}),  with respect to $\boldsymbol{\beta}$.  With this purpose, we need the following derivatives,
\begin{align*}
\frac{\partial \boldsymbol{\pi}^T_{i}(\boldsymbol{\beta})}{\partial \boldsymbol{\beta}} &= \boldsymbol{\Delta}^*(\boldsymbol{\pi}_{i}(\boldsymbol{\beta})) \otimes \boldsymbol{x}_i,\\
\frac{\partial}{\partial \boldsymbol{\beta}} \sum_{j=1}^{d+1} \pi_{ij}^{\alpha}(\boldsymbol{\beta})y_{ij}&=\frac{\partial}{\partial \boldsymbol{\beta}}\left\{\boldsymbol{1}^T_{d+1}  \text{diag}^{\alpha}(\boldsymbol{\pi}_i(\boldsymbol{\beta}))\boldsymbol{y}_i  \right\}\\
&= \alpha \ \boldsymbol{\Delta}^*(\boldsymbol{\pi}_i(\boldsymbol{\beta})) \text{diag}^{\alpha-1}(\boldsymbol{\pi}_i(\boldsymbol{\beta}))\boldsymbol{y}_i\otimes \boldsymbol{x}_i,
\\
\frac{\partial}{\partial \boldsymbol{\beta}} \sum_{j=1}^{d+1} \pi_{ij}^{\alpha+1}(\boldsymbol{\beta})&=\frac{\partial}{\partial \boldsymbol{\beta}}\left\{\boldsymbol{1}^T_{d+1}  \text{diag}^{\alpha+1}(\boldsymbol{\pi}_i(\boldsymbol{\beta}))\boldsymbol{1}_{d+1} \right\}\\
&= (\alpha+1) \ \boldsymbol{\Delta}^*(\boldsymbol{\pi}_i(\boldsymbol{\beta})) \text{diag}^{\alpha}(\boldsymbol{\pi}_i(\boldsymbol{\beta}))\boldsymbol{1}_{d+1}\otimes \boldsymbol{x}_i\\
&= (\alpha+1) \ \boldsymbol{\Delta}^*(\boldsymbol{\pi}_i(\boldsymbol{\beta})) \text{diag}^{\alpha-1}(\boldsymbol{\pi}_i(\boldsymbol{\beta}))\boldsymbol{\pi}_i(\boldsymbol{\beta})\otimes \boldsymbol{x}_i,
\end{align*}
being $\boldsymbol{1}_{d+1}$ the column vector of dimension $d+1$ with all entries being one.

\begin{theorem}\label{th:estimating}
The minimum RP estimator with tuning parameter $\alpha$ is obtained by solving the following system of equations
\begin{align}\label{eq:estRP}
\sum_{i=1}^n \boldsymbol{\Psi}_{i,\alpha}(\boldsymbol{\beta})=\boldsymbol{0}_{d(k+1)},
\end{align}
where 
\begin{align*}
\boldsymbol{\Psi}_{i,\alpha}(\boldsymbol{\beta})=\frac{1}{\left(\Gamma_i(\alpha,\boldsymbol{\beta})\right)^{\frac{\alpha}{1+\alpha}}}\boldsymbol{\Delta}^*(\boldsymbol{\pi}_i(\boldsymbol{\beta})) \text{diag}^{\alpha-1}\left(\boldsymbol{\pi}_i(\boldsymbol{\beta})\right)\left[\boldsymbol{y}_i-\frac{\Upsilon_i(\alpha,\boldsymbol{\beta})}{\Gamma_i(\alpha,\boldsymbol{\beta})}\boldsymbol{\pi}_i(\boldsymbol{\beta})\right]\otimes \boldsymbol{x}_i,
\end{align*}
and
\begin{align}
\Gamma_i(\alpha,\boldsymbol{\beta})&= \sum_{j=1}^{d+1} \pi_{ij}^{\alpha+1}(\boldsymbol{\beta})=\boldsymbol{1}^T_{d+1}  \text{diag}^{\alpha+1}(\boldsymbol{\pi}_i(\boldsymbol{\beta}))\boldsymbol{1}_{d+1},  \label{eq:Gamma}\\
\Upsilon_i(\alpha,\boldsymbol{\beta})&= \sum_{j=1}^{d+1} \pi_{ij}^{\alpha}(\boldsymbol{\beta})y_{ij}=\boldsymbol{1}^T_{d+1}  \text{diag}^{\alpha}(\boldsymbol{\pi}_i(\boldsymbol{\beta}))\boldsymbol{y}_{i}. \notag
\end{align}
In particular, for $\alpha=0$ , $\Gamma_i(0,\boldsymbol{\beta})=\Upsilon_i(0,\boldsymbol{\beta})=1$ for $i=1,\dots,n$, and the system of estimating equations in (\ref{eq:estRP}) coincides with tthe system of estimating equations  of the MLE (\ref{eq:estimatingMLE}).
\end{theorem}

\subsection{Asymptotic distribution}

In Castilla et al. (2022a, Section III) it was established that the asymptotic distribution of minimum RP estimators under the i.n.i.d.o setup is given by
\begin{align*}
\sqrt{n}\ \boldsymbol{\Omega}^{-1/2}_{n,\alpha}(\boldsymbol{\beta}^0)\boldsymbol{\Phi}_{n,\alpha}(\boldsymbol{\beta}^0)\left(\widehat{\boldsymbol{\beta}}_{\alpha}-\boldsymbol{\beta}^0\right)\overset{\mathcal{L}}{\underset{n\rightarrow \infty }{\longrightarrow }}\mathcal{N}\left( \boldsymbol{0}_{d(k+1)},\boldsymbol{I}_{d(k+1)}\right),
\end{align*}
where
\begin{align*}
\boldsymbol{\Phi}_{n,\alpha}(\boldsymbol{\beta}^0)&=\frac{1}{n}\sum_{i=1}^n \left(-\text{E}_{\boldsymbol{\beta}^0}\left[\frac{\partial^2 s_i^{(\alpha)}\left(\boldsymbol{Y},\boldsymbol{\beta} \right)}{\partial \boldsymbol{\beta}\partial  \boldsymbol{\beta}^T} \right]\right),\\
\boldsymbol{\Omega}_{n,\alpha}(\boldsymbol{\beta}^0)&=\frac{1}{n}\sum_{i=1}^n \left(\text{Var}_{\boldsymbol{\beta}^0}\left[ \frac{\partial s_i^{(\alpha)}\left(\boldsymbol{Y},\boldsymbol{\beta} \right)}{\partial \boldsymbol{\beta}}\right]\right).
\end{align*}

To particularize to the case of the PLRM, let us introduce some notation,
\begin{align}
\boldsymbol{\xi}_i(\alpha,\boldsymbol{\beta})&= \boldsymbol{\Delta}^*(\boldsymbol{\pi}_i(\boldsymbol{\beta})) \text{diag}^{\alpha-1}(\boldsymbol{\pi}_i(\boldsymbol{\beta}))\boldsymbol{\pi}_i(\boldsymbol{\beta})\otimes \boldsymbol{x}_i, \label{eq:xi}\\
\boldsymbol{J}_i(\alpha,\boldsymbol{\beta})&=\boldsymbol{\Delta}^*(\boldsymbol{\pi}_i(\boldsymbol{\beta}))\text{diag}^{\alpha-1}(\boldsymbol{\pi}_i(\boldsymbol{\beta}))\boldsymbol{\Delta}^{*T}(\boldsymbol{\pi}_i(\boldsymbol{\beta})) \otimes \boldsymbol{x}_i \boldsymbol{x}_i^T \label{eq:J}.
\end{align}
After some  algebraic computations, the asymptotic distribution is derived in the following Theorem.

\begin{theorem}\label{th:asym}
 Given the PLRM in (\ref{eq:PLRM}), the asymptotic distribution of the minimum RP estimator is given by
\begin{align}
\sqrt{n}\left(\widehat{\boldsymbol{\beta}}_{\alpha}-\boldsymbol{\beta}^0\right)\overset{\mathcal{L}}{\underset{n\rightarrow \infty }{\longrightarrow }}\mathcal{N}\left( \boldsymbol{0}_{d(k+1)},\boldsymbol{\Phi}^{-1}_{\alpha}(\boldsymbol{\beta}^0)\boldsymbol{\Omega}_{\alpha}(\boldsymbol{\beta}^0)\boldsymbol{\Phi}^{-1}_{\alpha}(\boldsymbol{\beta}^0)\right),
\end{align}
where $\boldsymbol{\beta}^0$ denotes the true value of  $\boldsymbol{\beta}$ and 
\begin{align*}
\boldsymbol{\Phi}_{\alpha}(\boldsymbol{\beta})&=\lim_{n\rightarrow \infty}\boldsymbol{\Phi}_{n,\alpha}(\boldsymbol{\beta})= \lim_{n\rightarrow \infty} \frac{1}{n}\sum_{i=1}^n \frac{1}{\left(\Gamma_i(\alpha,\boldsymbol{\beta})\right)^{\frac{\alpha}{1+\alpha}}}\left[\boldsymbol{J}_i(\alpha,\boldsymbol{\beta})-\frac{1}{\Gamma_i(\alpha,\boldsymbol{\beta})}\boldsymbol{\xi}_i(\alpha,\boldsymbol{\beta})\boldsymbol{\xi}^T_i(\alpha,\boldsymbol{\beta}) \right],\\
\boldsymbol{\Omega}_{\alpha}(\boldsymbol{\beta})&=\lim_{n\rightarrow \infty}\boldsymbol{\Omega}_{n,\alpha}(\boldsymbol{\beta})\\
&=\lim_{n\rightarrow \infty}\frac{1}{n}\sum_{i=1}^n \frac{1}{\left(\Gamma_i(\alpha,\boldsymbol{\beta})\right)^{\frac{2\alpha}{1+\alpha}}}\boldsymbol{J}_i(2\alpha,\boldsymbol{\beta})\\
& \quad  +\frac{1}{\left(\Gamma_i(\alpha,\boldsymbol{\beta})\right)^{1+\frac{2\alpha}{1+\alpha}}}\left[\frac{\Gamma_i(2\alpha,\boldsymbol{\beta})}{\Gamma_i(\alpha,\boldsymbol{\beta})} \boldsymbol{\xi}_i(\alpha,\boldsymbol{\beta})\boldsymbol{\xi}_i^T(\alpha,\boldsymbol{\beta})-\boldsymbol{\xi}_i(2\alpha,\boldsymbol{\beta})\boldsymbol{\xi}_i^T(\alpha,\boldsymbol{\beta})-\boldsymbol{\xi}_i(\alpha,\boldsymbol{\beta})\boldsymbol{\xi}_i^T(2\alpha,\boldsymbol{\beta})\right],
\end{align*}
with $\Gamma_i(\alpha,\boldsymbol{\beta})$, $\boldsymbol{\xi}_i(\alpha,\boldsymbol{\beta})$ and $\boldsymbol{J}_i(\alpha,\boldsymbol{\beta})$ defined in (\ref{eq:Gamma}), (\ref{eq:xi}) and (\ref{eq:J}), respectively. In particular, for $\alpha=0$, the asymptotic distribution is simplified to the asymptotic distribution of the MLE (\ref{eq:asymMLE}).
\end{theorem}


In practice, it is not possible to compute $\boldsymbol{\Phi}_{\alpha}(\boldsymbol{\beta}^0)$ and $\boldsymbol{\Omega}_{\alpha}(\boldsymbol{\beta}^0)$ as we do not know the exact value of our parameter vector, $\boldsymbol{\beta}^0$. Therefore, these matrices are empirically estimated by imputing the  minimum RP estimator, $\widehat{\boldsymbol{\beta}}_{\alpha}$.



\section{Wald-type tests for testing linear hypotheses \label{sec:Wald}}

Let us assume that we want to test a linear hypothesis involving our parameter vector $\boldsymbol{\beta}$. This can be expressed on the form 
\begin{align}\label{eq:H0}
H_0: \boldsymbol{L}\boldsymbol{\beta}=\boldsymbol{l} \quad \text{ against } \quad H_1: \boldsymbol{L}\boldsymbol{\beta}\neq \boldsymbol{l},
\end{align}
where $\boldsymbol{L}$ is a full-rank matrix with $d(k+1)$ columns and $r\leq (k+1)d$ rows and $\boldsymbol{l}$ is a vector of dimension $r$.
 
\begin{example}
Let us assume that we have a response variable with $d+1=3$ categories and $k=1$ explanatory variable. In this case, the parameter vector takes the form $\boldsymbol{\beta}=(\beta_{01},\beta_{11},\beta_{02},\beta_{12})^T$. The linear hypothesis $H_0:\beta_{11}-\beta_{12}=0$ can be expressed as in (\ref{eq:H0}) by setting $\boldsymbol{L}=(0,1,0,-1)$ and $\boldsymbol{l}=0$. Here, $r=1$, which corresponds to the number of independent linear relationships in the null hypothesis
\end{example}

Let us define the following matrix
\begin{align*}
\boldsymbol{V}_{n,\alpha}(\widehat{\boldsymbol{\beta}}_{\alpha})=\boldsymbol{\Phi}^{-1}_{n,\alpha}(\widehat{\boldsymbol{\beta}}_{\alpha})\boldsymbol{\Omega}_{n,\alpha}(\widehat{\boldsymbol{\beta}}_{\alpha})\boldsymbol{\Phi}^{-1}_{n,\alpha}(\widehat{\boldsymbol{\beta}}_{\alpha}).
\end{align*}

\begin{definition}
Let $\widehat{\boldsymbol{\beta}}_{\alpha}$ be the minimum RP estimator of $\boldsymbol{\beta} $ in the PLRM defined in (\ref{eq:PLRM}). Then, the family of Wald-type test statistics for testing the null hypothesis given in (\ref{eq:H0}) is defined as
\begin{align}\label{eq:Wald}
W_{n,\alpha}(\widehat{\boldsymbol{\beta}}_{\alpha})=n\left(\boldsymbol{L} \widehat{\boldsymbol{\beta}}_{\alpha}-\boldsymbol{l}\right)^T\left[\boldsymbol{L}\boldsymbol{V}_{n,\alpha}(\widehat{\boldsymbol{\beta}}_{\alpha})\boldsymbol{L}^T \right]^{-1}\left(\boldsymbol{L} \widehat{\boldsymbol{\beta}}_{\alpha}-\boldsymbol{l}\right).
\end{align}
In particular, for $\alpha=0$, $\boldsymbol{V}_{n,0}(\widehat{\boldsymbol{\beta}}_{0})=\boldsymbol{\Omega}^{-1}_{n}(\widehat{\boldsymbol{\beta}}_{\text{MLE}})$ and $W_{n,\alpha}(\widehat{\boldsymbol{\beta}}_{\alpha})$ becomes the classical MLE-based Wald test.
\end{definition}

By Theorem \ref{th:asym} we know that,
\begin{align*}
\sqrt{n}\left(\widehat{\boldsymbol{\beta}}_{\alpha}-\boldsymbol{\beta}^0\right)\overset{\mathcal{L}}{\underset{n\rightarrow \infty }{\longrightarrow }}\mathcal{N}\left( \boldsymbol{0}_{d(k+1)},\boldsymbol{V}_{\alpha}(\boldsymbol{\beta}^0)\right),
\end{align*}
where $ \boldsymbol{V}_{\alpha}(\boldsymbol{\beta}^0)=\lim_{n\rightarrow \infty} \boldsymbol{V}_{n,\alpha}(\boldsymbol{\beta}^0)$. Therefore, 
\begin{align*}
\sqrt{n}\left(\boldsymbol{L} \boldsymbol{\beta}^0-\boldsymbol{l}\right)\overset{\mathcal{L}}{\underset{n\rightarrow \infty }{\longrightarrow }}\mathcal{N}\left( \boldsymbol{0}_{r},\boldsymbol{L} \boldsymbol{V}_{\alpha}(\boldsymbol{\beta}^0)\boldsymbol{L}^T\right).
\end{align*}
As $rank(\boldsymbol{L})=r$, we have the  following result  taking into account that $\widehat{\boldsymbol{\beta}}_{\alpha}$ is a consistent estimator of $\boldsymbol{\beta}^0$.

\begin{theorem}
The asymptotic distribution of the Wald-type test statistics, $W_{n,\alpha}(\widehat{\boldsymbol{\beta}}_{\alpha})$, under the null hypothesis in (\ref{eq:H0}) is a chi-square distribution with $r$ degrees of freedom, $\chi_{r}^2$.
\end{theorem}

Based on the previous theorem, the null hypothesis in (\ref{eq:H0}) will be rejected if
$$
W_{n,\alpha}(\widehat{\boldsymbol{\beta}}_{\alpha})>\chi^2_{r,\tau},
$$
being $\chi^2_{r,\tau}$ the $100(1 - \tau)$ percentile of a chi-square distribution with $r$ degrees of freedom.\\

Now, we present some results pertaining to the Wald-type tests, the proofs of which are detailed in Appendix \ref{app:proofsWald}.
\begin{theorem}\label{th:WaldConsistent}
Given $\boldsymbol{\beta}^1$ verifying $\boldsymbol{L}^T \boldsymbol{\beta}^1 \neq \boldsymbol{l}$, then the Wald type-test in (\ref{eq:Wald}) is consistent in the sense of Fraser (1957), i.e., 
\begin{align*}
\lim_{n\rightarrow \infty} P_{\boldsymbol{\beta}^1}\left(W_{n,\alpha}(\widehat{\boldsymbol{\beta}}_{\alpha})>\chi^2_{r,\tau} \right)=1.
\end{align*}
\end{theorem}

\begin{proposition}\label{th:PowerSize}
The necessary sample size $N_{\Pi_0}$ for the Wald-type tests to have a predetermined power, $\Pi_0$,
is given by 
\begin{align*}
N_{\Pi_0}=1+\left[\frac{A_1+A_2+\sqrt{A_1(A_1+2A_2)}}{2 l^2_{\boldsymbol{\beta}^1}(\boldsymbol{\beta}^1)} \right],
\end{align*}
being $[\cdot]$ the integer part, 
\begin{align*}
A_1&=\left.\frac{\partial l_{\widehat{\boldsymbol{\beta}}_{\alpha}}(\boldsymbol{\beta})}{\partial \boldsymbol{\beta}^T} \right|_{\boldsymbol{\beta}=\boldsymbol{\beta}^1}
\boldsymbol{V}_{\alpha}(\boldsymbol{\beta}^1)\left.\frac{\partial l_{\widehat{\boldsymbol{\beta}}_{\alpha}}(\boldsymbol{\beta})}{\partial \boldsymbol{\beta}} \right|_{\boldsymbol{\beta}=\boldsymbol{\beta}^1}\left( \phi^{-1}(1-\Pi_0)\right)^{2},\\
 A_2&=2 l_{\boldsymbol{\beta}^1}(\boldsymbol{\beta}^1) \chi^2_{r,\tau},
\end{align*}
with $\phi(\cdot)$  the distribution function of a standard normal distribution, and
\begin{align*}
l_{\boldsymbol{\beta}^{\ast}}(\boldsymbol{\beta}^{\ast \ast})=\left(\boldsymbol{L}\boldsymbol{\beta}^{\ast}-\boldsymbol{l} \right)^T \left(\boldsymbol{L} \boldsymbol{V}_{\alpha}(\boldsymbol{\beta}^{\ast \ast})\boldsymbol{L}^T\right)^{-1}\left(\boldsymbol{L}\boldsymbol{\beta}^{\ast}-\boldsymbol{l} \right).
\end{align*}
\end{proposition}


\section{Influence functions of minimum RP estimators and Wald-type tests \label{sec:IF}} 

Introduced by Hampel (1986), the influence function serves as a measure of the standardized asymptotic bias (approximated to the first order) that results from an infinitesimal contamination at a given point $t$. In particular,  for any estimator defined in terms of a statistical functional $T(G)$ from the true distribution function $G$, its influence function is given by
 $$
 IF(t,G)=\lim_{\varepsilon \downarrow 0}\frac{T(G_{\varepsilon})-T(G)}{\varepsilon}=\left. \frac{\partial T(G_{\varepsilon})}{\partial \varepsilon}\right|_{\varepsilon=0^`+},
 $$
with $G_{\varepsilon}= (1 -\varepsilon)G + \varepsilon \Delta_t$,  $\varepsilon$ being the contamination proportion and $\Delta_t$ being the degenerate distribution at $t$. The maximum of this influence function over $t$ indicates the extent of bias due to contamination, and thus, the smaller its value, the more robust the estimator may be. While this is an important concept that has been widely studied in the literature, it's also crucial to note that the influence function does not always capture the robustness of a particular statistic. For instance, Cressie-Read divergences with a negative tuning parameter, with the well-known robust Hellinger distance as a specific case, are demonstrated to have the same unbounded influence function as the MLE in many statistical contexts. See Lindsay (1994), Basu et al. (1997), and Castilla and Chocano (2022). In this study, we examine the influence functions of minimum RP estimators and derived Wald-type tests under the PLRM.

In the previous sections, we have seen that under the PLRM,  the minimum RP estimator with tuning parameter $\alpha$ of $\boldsymbol{\beta}$, $\widehat{\boldsymbol{\beta}}_{\alpha}$, is given by (\ref{eq:Halpha}) or, equivalently by 
\begin{align*}
\sum_{i=1}^n \boldsymbol{\Psi}_{i,\alpha}(\boldsymbol{\beta})=\boldsymbol{0}_{d(k+1)},
\end{align*}
where 
\begin{align*}
\boldsymbol{\Psi}_{i,\alpha}(\boldsymbol{\beta})=\frac{1}{\left(\Gamma_i(\alpha,\boldsymbol{\beta})\right)^{\frac{\alpha}{1+\alpha}}}\boldsymbol{\Delta}^*(\boldsymbol{\pi}_i(\boldsymbol{\beta})) \text{diag}^{\alpha-1}\left(\boldsymbol{\pi}_i(\boldsymbol{\beta})\right)\left[\boldsymbol{y}_i-\frac{\left[\boldsymbol{1}^T_{d+1}  \text{diag}^{\alpha}(\boldsymbol{\pi}_i(\boldsymbol{\beta}))\boldsymbol{y}_{i}\right]}{\Gamma_i(\alpha,\boldsymbol{\beta})}\boldsymbol{\pi}_i(\boldsymbol{\beta})\right]\otimes \boldsymbol{x}_i.
\end{align*}
For convenience, let us denote 
$$
\boldsymbol{D}_{i,\alpha}(\boldsymbol{\beta})=\frac{1}{\left(\Gamma_i(\alpha,\boldsymbol{\beta})\right)^{\frac{\alpha}{1+\alpha}}}\boldsymbol{\Delta}^*(\boldsymbol{\pi}_i(\boldsymbol{\beta})) \text{diag}^{\alpha-1}\left(\boldsymbol{\pi}_i(\boldsymbol{\beta})\right),
$$
so then

\begin{align*}
\boldsymbol{\Psi}_{i,\alpha}(\boldsymbol{\beta})=\boldsymbol{D}_{i,\alpha}(\boldsymbol{\beta}) \left[\boldsymbol{y}_i-\frac{\left[\boldsymbol{1}^T_{d+1}  \text{diag}^{\alpha}(\boldsymbol{\pi}_i(\boldsymbol{\beta}))\boldsymbol{y}_{i}\right]}{\Gamma_i(\alpha,\boldsymbol{\beta})}\boldsymbol{\pi}_i(\boldsymbol{\beta})\right]\otimes \boldsymbol{x}_i.
\end{align*}
If we denote $T_{\alpha}$ the functional associated with the minimum RP estimator, then when the assumption of PLRM holds,   $\boldsymbol{\beta}^0=T_{\alpha}(\boldsymbol{G})$, where $\boldsymbol{G}=(G_1,\dots,G_n)^T$ is the true distribution function.

Note that, in such non-homogeneous settings, outliers can be either in any one or more index $i\in\{1,\dots,n\}$. Let $G_{i,\varepsilon}=(1-\varepsilon)G_i+\varepsilon \Delta_i$. Following this notation,
$$\boldsymbol{\beta}^{i_0}_{\alpha,\varepsilon}=T_{\alpha}(\boldsymbol{G}_{\varepsilon}^{i_0})=T_{\alpha}(G_1,\dots,G_{i_0,\varepsilon},\dots,G_n)$$
is the minimum RP functional with contamination only in the $i_0$-th direction, and 
$$\boldsymbol{\beta}_{\alpha,\varepsilon}=T_{\alpha}(\boldsymbol{G}_{\varepsilon})=T_{\alpha}(G_{1,\varepsilon},\dots,G_{i,\varepsilon},\dots,G_{n,\varepsilon})$$
is the minimum RP functional with contamination in all directions. Now, following Theorem 10 in Castilla et al. (2022a), we have that the influence function when there is contamination in only one specific index $i_0$ at the point $\boldsymbol{t}_{i_0}$
is given by

\begin{align}
IF_{i_0}(\boldsymbol{t}_{i_0},T_{\alpha},\boldsymbol{\beta}^0)=\boldsymbol{\Phi}^{-1}_{n,\alpha}(\boldsymbol{\beta}^0)\boldsymbol{D}_{i_0,\alpha}(\boldsymbol{\beta}^0)\left[\boldsymbol{t}_{i_0}-\frac{\left[\boldsymbol{1}^T_{d+1}  \text{diag}^{\alpha}(\boldsymbol{\pi}_{i_0}(\boldsymbol{\beta}^0))\boldsymbol{t}_{i_0}\right]}{\Gamma_{i_0}(\alpha,\boldsymbol{\beta}^0)}\boldsymbol{\pi}_{i_0}(\boldsymbol{\beta}^0)\right]\otimes \boldsymbol{x}_{i_0}. 
\end{align}
When the contamination is all $n$ distributions $G_i$ at the points $\boldsymbol{t}_i$, $i=1,\dots,n$, then, denoting $\underline{\boldsymbol{t}}=(\boldsymbol{t}_1^T,\dots,\boldsymbol{t}_n^T)^T$, the influence function is given by
\begin{align}
IF(\underline{\boldsymbol{t}},T_{\alpha},\boldsymbol{\beta}^0)=\boldsymbol{\Phi}^{-1}_{n,\alpha}(\boldsymbol{\beta}^0)\sum_{i=1}^n\boldsymbol{D}_{i,\alpha}(\boldsymbol{\beta}^0) \left[\boldsymbol{t}_i-\frac{\left[\boldsymbol{1}^T_{d+1}  \text{diag}^{\alpha}(\boldsymbol{\pi}_i(\boldsymbol{\beta}^0))\boldsymbol{t}_{i}\right]}{\Gamma_i(\alpha,\boldsymbol{\beta}^0)}\boldsymbol{\pi}_i(\boldsymbol{\beta}^0)\right]\otimes \boldsymbol{x}_i.
\end{align}

Further, the second-order influence functions associated to the Wald-type tests in (\ref{eq:Wald}) (with associated statistical functional $W_{\alpha}$) are given by
\begin{align*}
IF_{i_0}(\boldsymbol{t}_{i_0},W_{\alpha},\boldsymbol{\beta}^0)&=2 IF^T_{i_0}(\boldsymbol{t}_{i_0},T_{\alpha},\boldsymbol{\beta}^0) \left[ \boldsymbol{L}^T\left[\boldsymbol{L}\boldsymbol{V}_{n,\alpha}(\boldsymbol{\beta}^0)\boldsymbol{L}^T \right]^{-1}\boldsymbol{L}\right] IF_{i_0}(\boldsymbol{t}_{i_0},T_{\alpha},\boldsymbol{\beta}^0),\\ 
IF(\underline{\boldsymbol{t}},W_{\alpha},\boldsymbol{\beta}_0)&=2 IF^T(\underline{\boldsymbol{t}},T_{\alpha},\boldsymbol{\beta}^0)\left[ \boldsymbol{L}^T\left[\boldsymbol{L}\boldsymbol{V}_{n,\alpha}(\boldsymbol{\beta}^0)\boldsymbol{L}^T \right]^{-1}\boldsymbol{L}\right] IF(\underline{\boldsymbol{t}},T_{\alpha},\boldsymbol{\beta}^0).
\end{align*}
Therefore, the boundedness of these influence functions is directly dependent on the boundedness of the minimum RP estimation.

Now, we can examine the boundedness of the obtained influence functions with respect to the response variable $\boldsymbol{y}$ or with respect to the predictors $\boldsymbol{x}$ (B-robustness). On one hand, it can be proved that the influence function of MLE is unbounded (Miron et al., 2022), while the influence function of minimum RP estimators is also unbounded for $k\geq 2$ (see Appendix \ref{app:Brobust} for a proof of this result). However, we cannot directly infer the robustness against outliers in the response variable, which are, in fact, the misspecification errors. This is because, in this case, $\boldsymbol{t}_i$ only changes its indicative category and the influence functions are bounded for all $\alpha \geq 0$. However, the lack of robustness of MLE in the PLRM against such misspecification errors is well-known. In the subsequent sections, we empirically prove that the minimum RP estimators (and Wald-type tests) are robust against this kind of errors.

\section{Monte Carlo simulation study \label{sec:MC}}

In this section, we carry out a comprehensive Monte Carlo simulation study to assess the performance of the proposed estimators and Wald-type tests for various values of the tuning parameter $\alpha$. All simulations presented here have been executed using the R software with $1,000$ replications.

Adopting an approach similar to that of Castilla et al. (2018), our simulated data consist of two explanatory variables ($k=2$) generated from the standard normal distribution and an explanatory variable with three response categories ($d+1=3$) generated under the PLRM with $\boldsymbol{\beta}^0=(0, -0.9,  0.1,  0.6, -1.2,  0.8)^T$. To evaluate the robustness of the proposed procedures, different percentages of responses are drawn from a multinomial distribution with interchanged first and third conditional probabilities. 

\begin{figure}[p]
\includegraphics[scale=0.555]{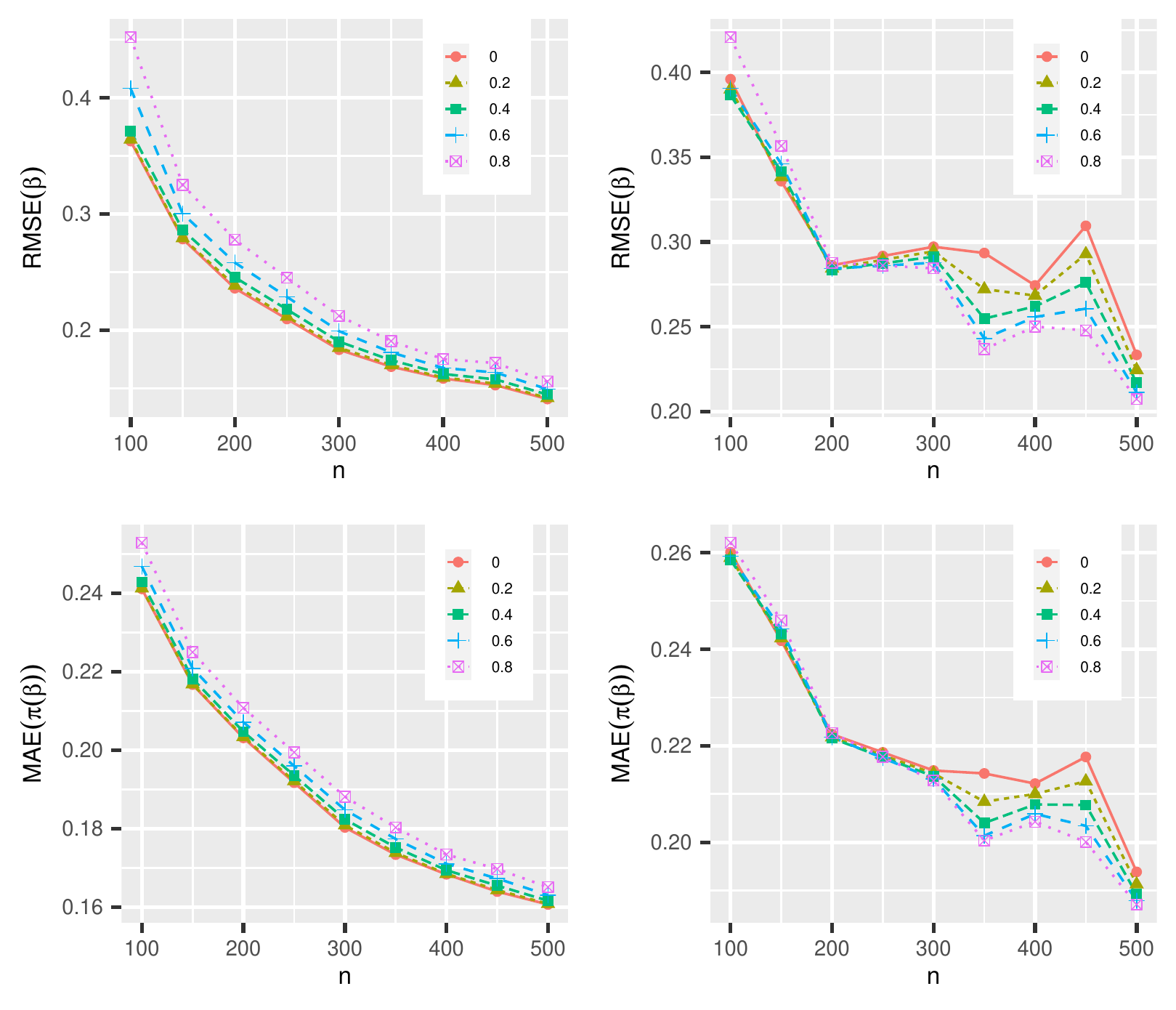}
\caption{RMSE of the model parameter vector and MAE of estimated probabilities, comparing pure data (left) and $10\%$-contaminated data (right) for different sample sizes.\label{fig:MC_est}}
\end{figure}

\begin{figure}[p]
\includegraphics[scale=0.555]{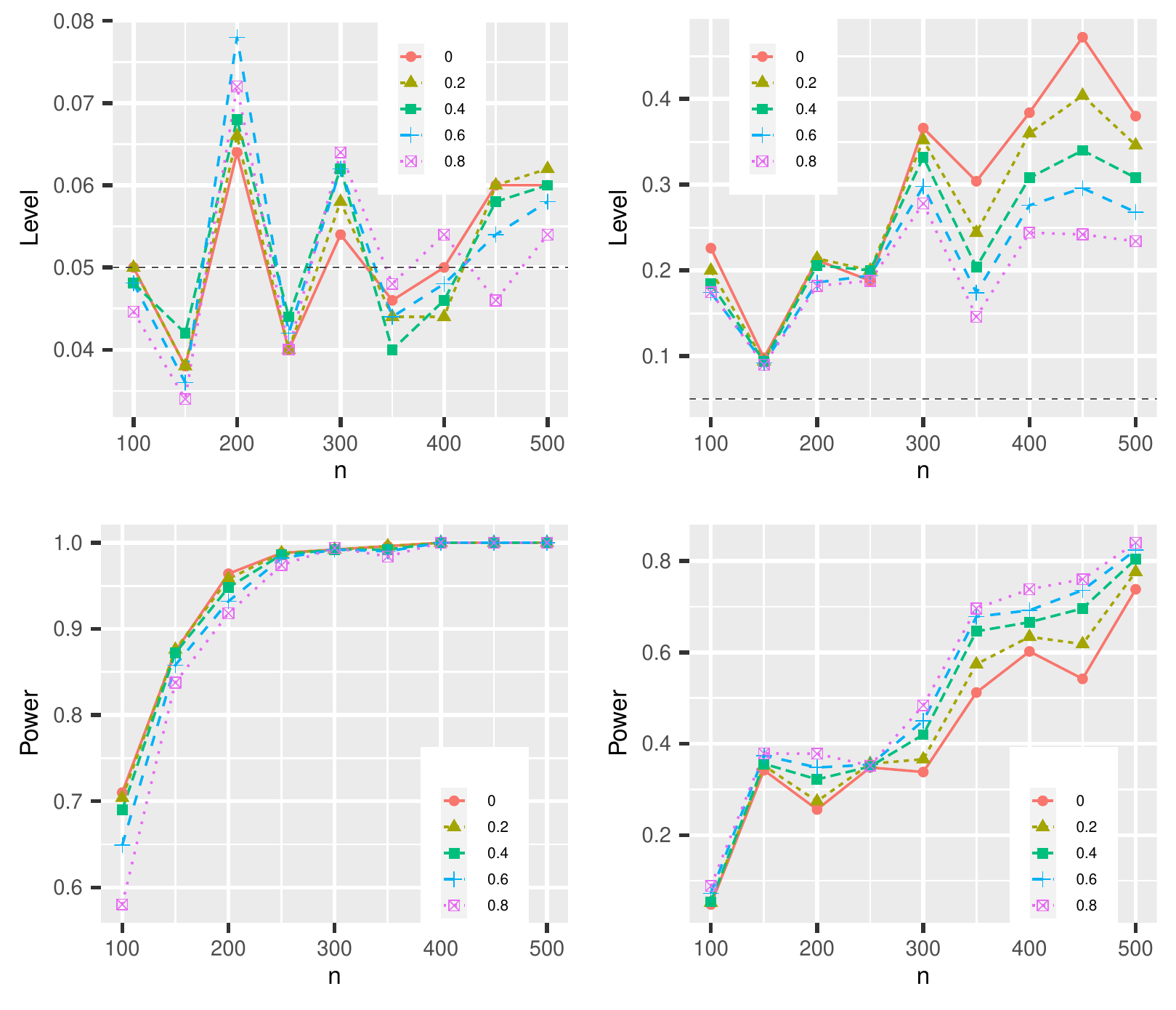}
\caption{Empirical levels and powers, comparing pure data (left) and $10\%$-contaminated data (right) for different sample sizes.\label{fig:MC_Wald}}
\end{figure}

\begin{figure}[p]
\includegraphics[scale=0.555]{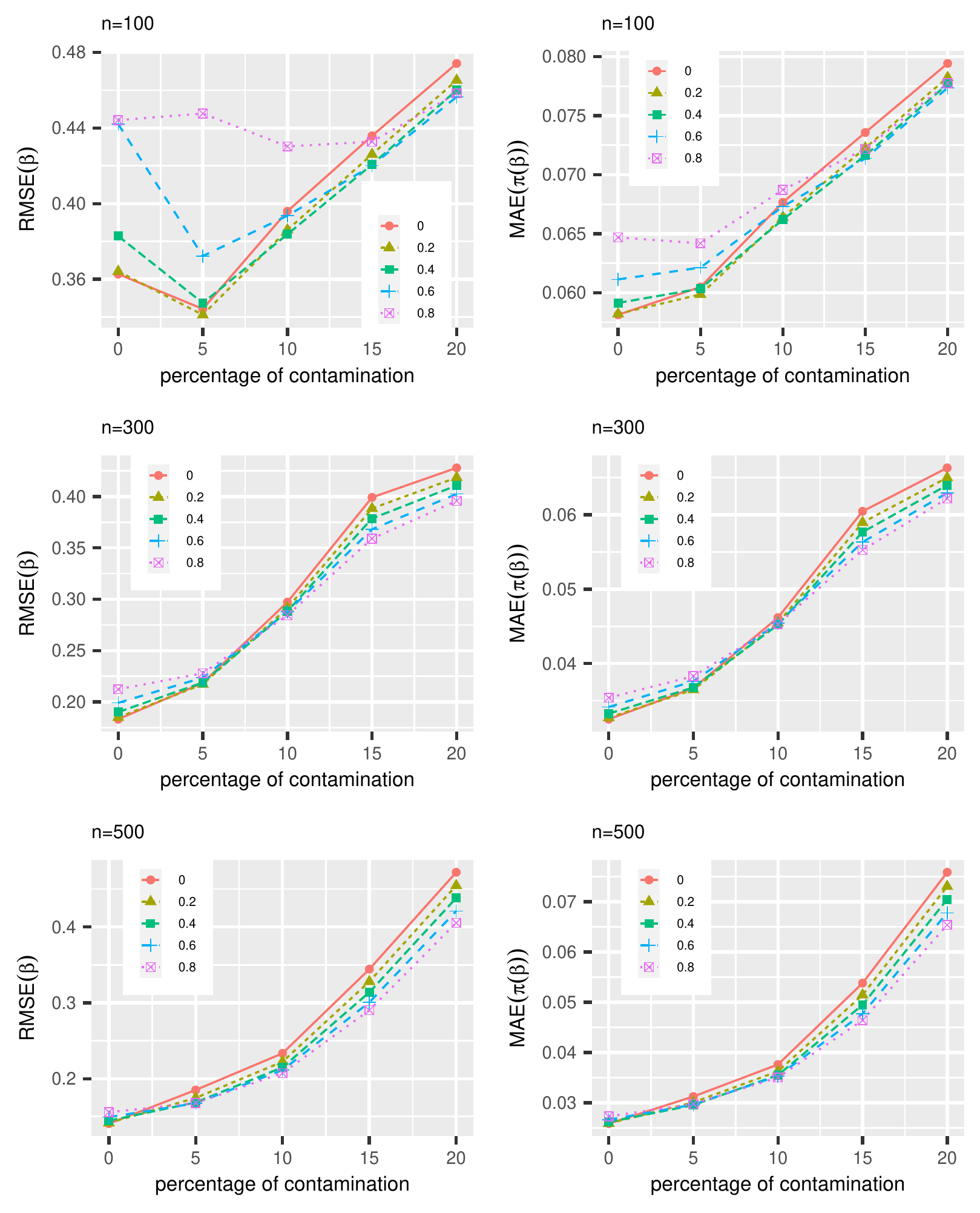}
\caption{RMSE of the model parameter vector (left) and MAE of estimated probabilities (right), presented for varying levels of contamination and sample sizes.\label{fig:MC_degree}}
\end{figure}

\subsection{Performance of the minimum RP estimators }

First of all, we evaluate the performance of the proposed estimators in terms of the root mean square error (RMSE) of the estimated vector of parameters $\widehat{\boldsymbol{\beta}}_{\alpha}$ and the mean absolute error (MAE) of the estimated probabilities $\boldsymbol{\pi}(\widehat{\boldsymbol{\beta}}_{\alpha})$ for different sample sizes $n\in \{100,150,\dots,500\}$, both for pure and  $10\%$-contaminated data (see Figure \ref{fig:MC_est}). To better evaluate the robustness of minimum RP estimators, the same simulation is carried our under different percentages of contamination $q\in \{0,5,10,15,20\}$ and different sample sizes (see Figure \ref{fig:MC_degree}).

As expected, the Maximum Likelihood Estimator (MLE) with $\alpha=0$ shows the highest efficiency when dealing with pure data, regardless of the sample size. It's worth noting, however, that the difference among minimum RP estimators with a small tuning parameter value is not substantial. An increase in either the sample size or the percentage of contaminated observations negatively impacts the performance of the MLE, revealing a significant lack of robustness. Minimum RP estimators are presented as a more robust alternative to MLE. However, a high value of the tuning parameter $\alpha$ may be detrimental in the absence of outliers.

\subsection{Performance of the Wald-type tests }

To study the performance of the proposed Wald-type tests, the null hypothesis $H_0:\beta_{02}=0.6$ is tested against the alternative hypothesis $H_0:\beta_{02} \neq 0.6$ for both pure and $10\%$ contaminate data under different sample sizes. The empirical levels, obtained as the proportion  of test statistics exceeding the corresponding chi-square critical value are presented in top of Figure  \ref{fig:MC_Wald}. In this case, we use a nominal level of $\tau=0.05$. The empirical powers, computed in a similar manner, under the alternative  hypothesis $H_1:\beta_{02} = 1.35$ are presented in the bottom of Figure   \ref{fig:MC_Wald}.

When pure data are considered, empirical levels of MLE and minimum RP estimators with  tuning parameter $\alpha>0$ are very close to the nominal level. The presence of outliers has a significantly negative impact on classical Wald-tests. This can be mitigated by employing Wald-type tests with $\alpha>0$. When dealing with pure data, classical Wald-tests exhibit the highest powers, but without a significant difference.  For contaminated data, Wald-type tests with $\alpha>0$ show superior performance for all sample sizes.

\begin{table}[t]\setlength{\tabcolsep}{10pt}\renewcommand{\arraystretch}{0.98}
\caption{ARE of minimum RP estimators for different samples sizes.\label{table:AREs}}
\center
\small
\begin{tabular}{|r|rrrrrr|}
  \hline
 $\alpha$ & $\beta_{01}$ & $\beta_{11}$ & $\beta_{21}$ & $\beta_{02}$ & $\beta_{12}$ & $\beta_{22}$ \\ 
  \hline
  $n=100$ & & & & & & \\
  0 & 1.0000 & 1.0000 & 1.0000 & 1.0000 & 1.0000 & 1.0000 \\ 
  0.2 & 0.9921 & 0.9875 & 0.9920 & 0.9899 & 0.9835 & 0.9907 \\ 
  0.4 & 0.9691 & 0.9539 & 0.9694 & 0.9620 & 0.9405 & 0.9648 \\ 
  0.6 & 0.9333 & 0.9054 & 0.9347 & 0.9209 & 0.8812 & 0.9255 \\ 
  0.8 & 0.8877 & 0.8480 & 0.8904 & 0.8710 & 0.8142 & 0.8757 \\ 
    \hline
  $n=200$ & & & & & & \\
  0 & 1.0000 & 1.0000 & 1.0000 & 1.0000 & 1.0000 & 1.0000 \\ 
  0.2 & 0.9923 & 0.9893 & 0.9888 & 0.9905 & 0.9873 & 0.9872 \\ 
  0.4 & 0.9689 & 0.9597 & 0.9588 & 0.9634 & 0.9529 & 0.9515 \\ 
  0.6 & 0.9319 & 0.9155 & 0.9164 & 0.9229 & 0.9033 & 0.8996 \\ 
  0.8 & 0.8846 & 0.8612 & 0.8671 & 0.8736 & 0.8442 & 0.8385 \\ 
   \hline
  $n=300$ & & & & & & \\
  0 & 1.0000 & 1.0000 & 1.0000 & 1.0000 & 1.0000 & 1.0000 \\ 
  0.2 & 0.9925 & 0.9874 & 0.9900 & 0.9905 & 0.9850 & 0.9883 \\ 
  0.4 & 0.9699 & 0.9531 & 0.9627 & 0.9637 & 0.9451 & 0.9568 \\ 
  0.6 & 0.9341 & 0.9027 & 0.9229 & 0.9238 & 0.8886 & 0.9112 \\ 
  0.8 & 0.8880 & 0.8421 & 0.8748 & 0.8755 & 0.8227 & 0.8568 \\ 
   \hline
\end{tabular}
\end{table}

\subsection{Effect on $\boldsymbol{\alpha}$: theory and practice \label{sec:ARE}}

In previous sections, we have empirically demonstrated that under a non-contaminated scenario in the PLRM, the MLE, as expected, yields the smallest error values. In this context, minimum RP estimators with a low to moderate value of the tuning parameter $\alpha>0$ do not show a significant decrease in efficiency. On the other hand, when dealing with contaminated data, minimum RP estimators with $\alpha>0$ exhibit substantially more robust behavior than the MLE.

Adopting a more theoretical approach, we now compare the asymptotic variances across different tuning parameters by computing the asymptotic relative efficiency (ARE), which is obtained as the ratio of their asymptotic variances with respect to the most efficient likelihood-based estimator at $\alpha=0$ (MLE). Results for the same simulation scheme considered throughout this section are presented in Table \ref{table:AREs}. As observed, although the ARE of minimum RP estimators with $\alpha>0$ is less than one, the loss in efficiency is not very high for small to moderate values of $\alpha$. This is in concordance with the empirical results obtained previously. Note also that, in practical applications, the computational process of estimation may encounter certain difficulties when dealing with elevated values of $\alpha$.  It seems therefore reasonable to choose a moderate value of $\alpha$, which may present a good trade-off between efficiency and robustness. However, developing an ad-hoc approach for choosing the optimal tuning parameter in each dataset could be more practical. Several methods have been discussed in the literature, with perhaps the most notable being the one proposed by Warwick and Jones (2005), and its iterated version by Basak et al. (2020). This method aims to minimize the estimated mean square error, which is computed as the sum of the squared (estimated) bias and variance in a range of potential tuning constants. If the range is too dense, the computation can become costly, and after conducting several numerical experiments, we found that this method was not particularly more effective for minimum RP estimators in the context of PLRM than simply choosing a tuning parameter with a moderate value of $\alpha$. Therefore, further research in this area could be beneficial for future work.


\section{Real data application: Diabetes dataset \label{sec:Data}}

Let us consider the Diabetes dataset presented in Section \ref{sec:small_example}. To evaluate the robustness of minimum RP estimators, we introduce contamination by randomly selecting $14$ observations (around $10\%$ of the total) and interchanging their category responses. We employ two distinct contamination schemes: In the first, the original \textit{normal}, \textit{chemical}, and \textit{overt} response categories are reassigned as the \textit{chemical}, \textit{overt}, and \textit{normal} categories, respectively. In the second scheme, they are reassigned as the \textit{overt}, \textit{normal}, and \textit{chemical} categories, respectively. These schemes serve to simulate the potential impact of misclassification or measurement error in real-world data. For each contamination scheme, we generate $200$ datasets, so our results may be more reliable than if we only considered the modified dataset presented in Section \ref{sec:small_example}, which will also be studied later for illustrative purposes. 

\begin{figure}[h!!t]
\center
\begin{tabular}{cc}
\includegraphics[scale=0.37]{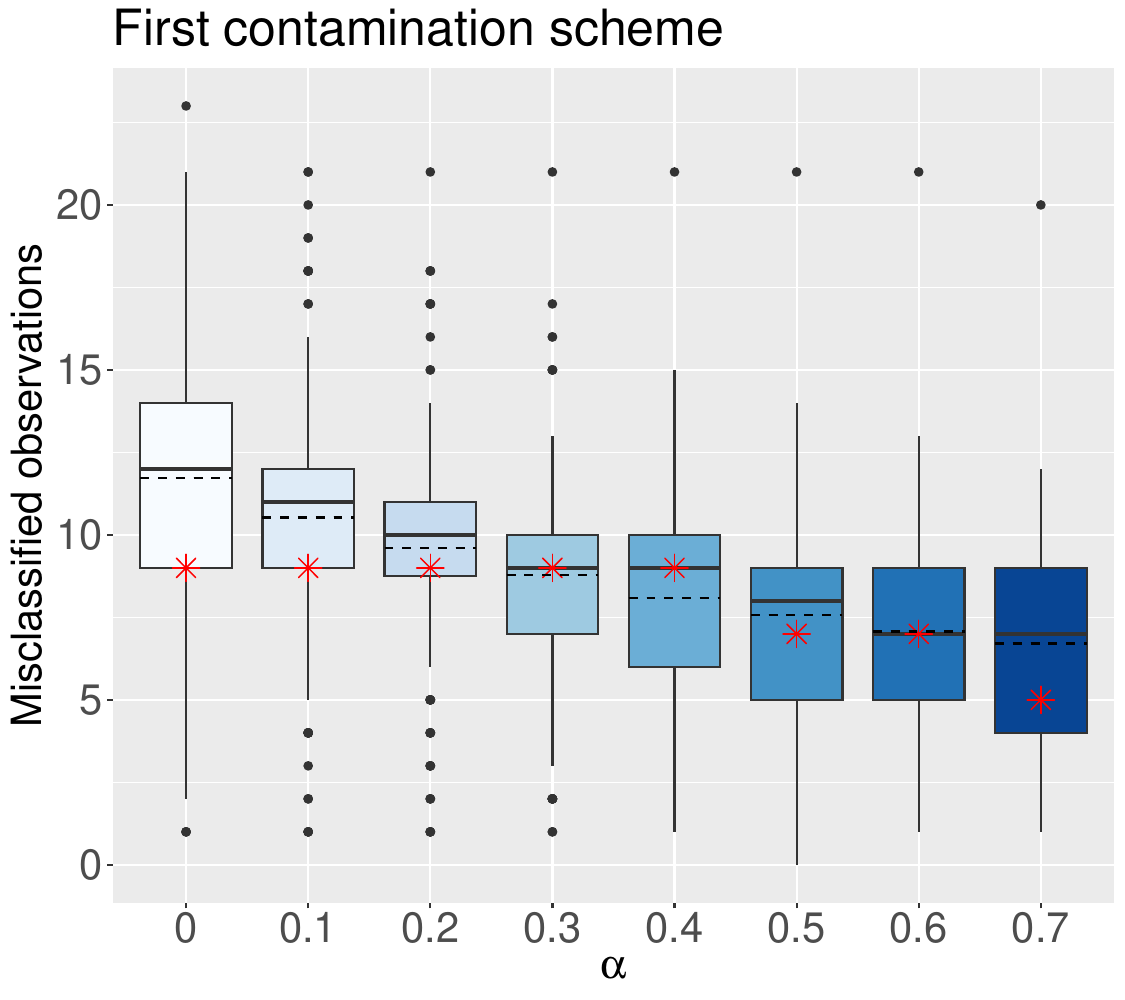}
 & \includegraphics[scale=0.37]{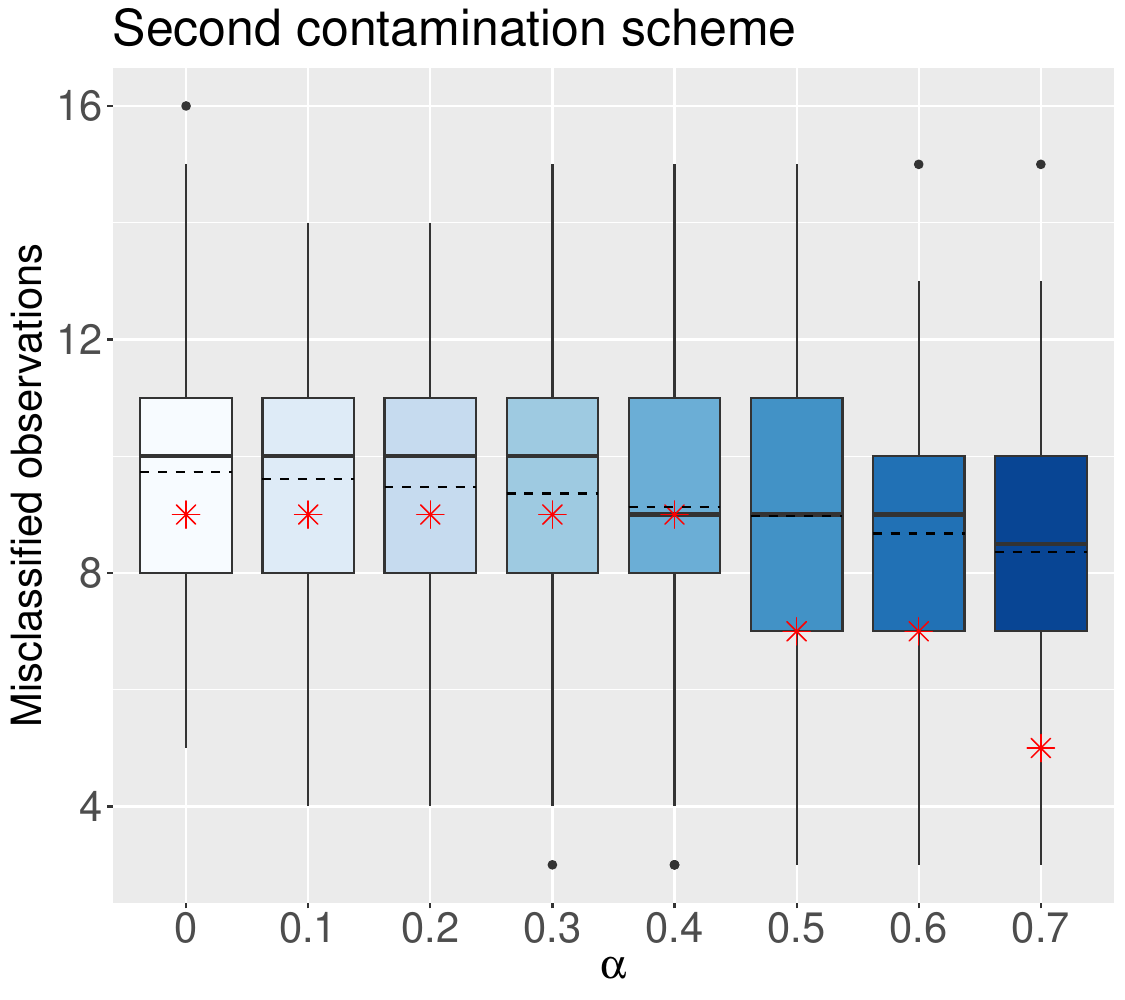}
\end{tabular} 
\caption{Diabetes Dataset: The plot quantifies misclassifications for various minimum RP estimators in  200 contaminated datasets. Red stars denote misclassifications in the unaltered dataset. \label{fig:DIABETES_boxs}}
\end{figure}

\begin{figure}[h!!t]
\center
\begin{tabular}{cc}
\includegraphics[scale=0.405]{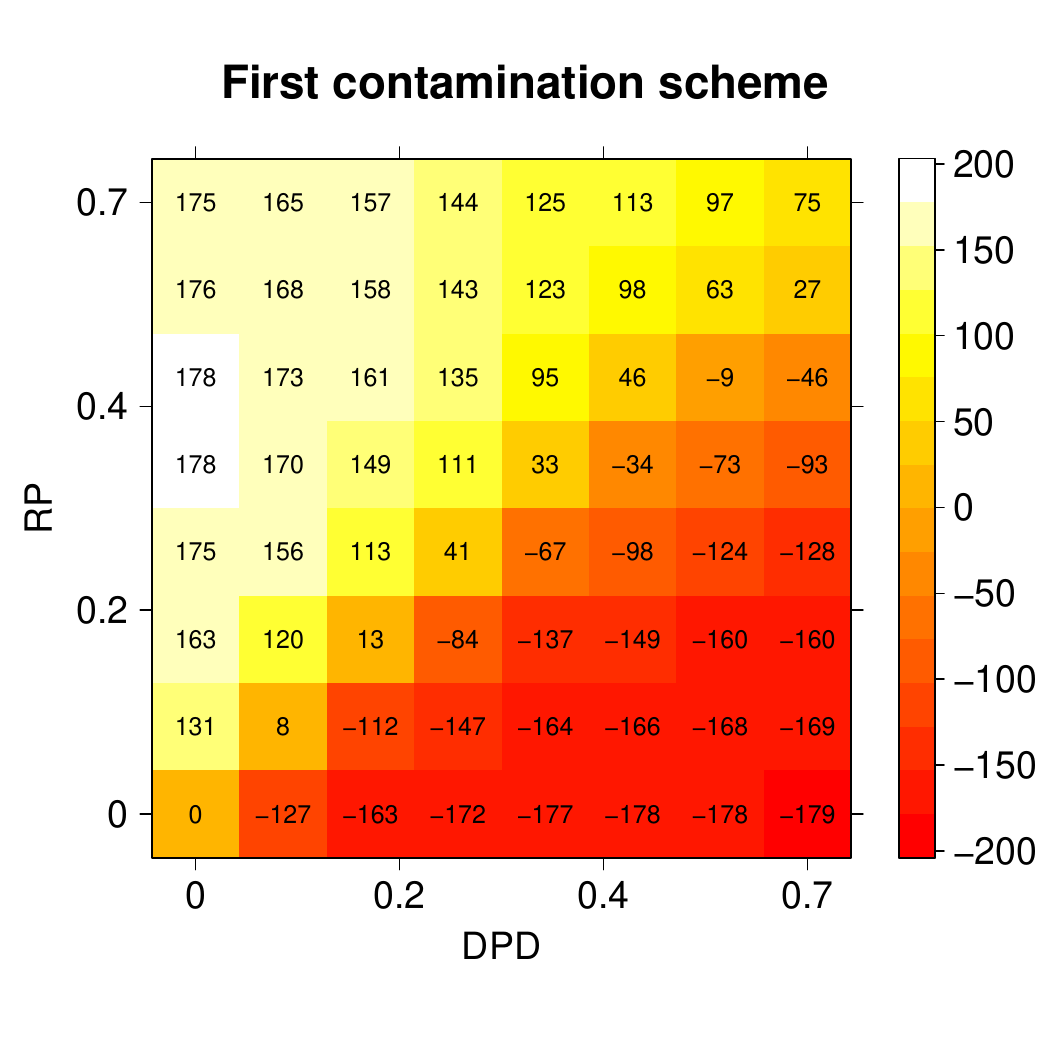}
 & \includegraphics[scale=0.405]{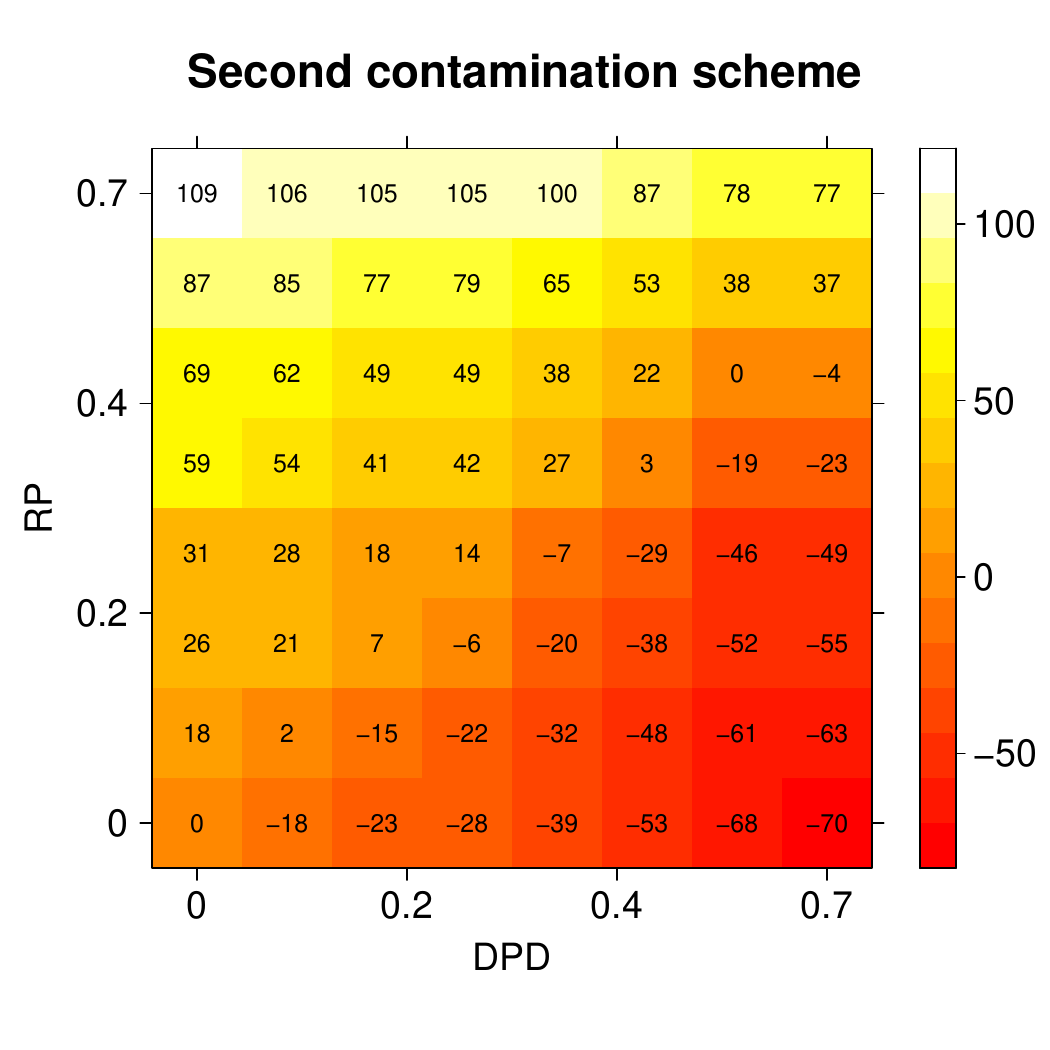}
\end{tabular} 
\vspace{-0.5cm}
\caption{Diabetes Dataset: Comparative Analysis of minimum RP and DPD estimators. The plot quantifies instances where RP estimators outperform DPD estimators in $200$ contaminated datasets. \label{fig:DIABETES_heat}}
\end{figure}

We compute minimum RP estimators for $\alpha\in[0,0.7]$, assign categories based on highest estimated probabilities, and compare these to the unaltered dataset. Differences, considered as misclassifications, are illustrated in Figure \ref{fig:DIABETES_boxs} for both contamination schemes. As expected, in most of the cases, the means of misclassifications (denoted by dashed lines in the boxplots) are higher than those for the unaltered dataset (denoted by red stars).  Misclassification values decrease with higher tuning parameters, a trend more evident in the first contamination scheme. These values and those of the example in Section \ref{sec:small_example}, are summarized in Table \ref{table:DiabetesCOMP}, for an easier comparison. Note that we are unable to compute minimum RP estimators with $\alpha> 0.7$, highlighting one of the main potential problems of employing high values of the tuning parameter. 

Errors in our example are higher than those of the contaminated schemes. This is probably because these misclassifications are concentrated in only one category, while the contaminated observations in the two schemes are randomly chosen among the total. When applying the Warwick and Jones procedure for choosing the optimal tuning parameter in a grid with a distance of $10^{-2}$ between potential candidates, as referred to in Section \ref{sec:ARE}, the proposed optimal tuning parameter for the original dataset is $\alpha=0$, with  $9$ misclassified observations.. For the modified dataset of our example, the optimal tuning parameter is $\alpha=0.4$, resulting in $21$ misclassified observations.

\subsubsection*{A comparison with minimum DPD estimators}

\begin{table}[t!!h!!]\setlength{\tabcolsep}{10pt}\renewcommand{\arraystretch}{0.98}
\caption{Diabetes Dataset: Comparative Analysis of Minimum RP and DPD Estimators. This table quantifies the number of misclassified observations in the original dataset and the modified dataset in Section  \ref{sec:small_example}, and the average number of misclassified responses in $200$ contaminated datasets.\label{table:DiabetesCOMP}}
\center
\small
\begin{tabular}{|l|r|rrrr|}
  \hline
 Method & Tuning   & \multicolumn{4}{c|}{Misclassifications} \\ 
   & parameter &  original data & example&   first scheme & second  scheme\\
  \hline
MLE       & $0$ & 9    &29&11.72 & 9.73 \\  \hline
RP & $\alpha=0.10$ & 9 &25& 10.53 & 9.61 \\ 
    & $\alpha=0.20$ & 9 &23& 9.60 & 9.47 \\ 
    & $\alpha=0.30$ & 9 &23& 8.78 & 9.36 \\ 
    & $\alpha=0.40$ & 9 &21& 8.10 & 9.12 \\ 
    & $\alpha=0.50$ & 7 &20& 7.57 & 8.97 \\ 
    & $\alpha=0.60$ & 7 &19&  7.07 & 8.68 \\ 
    & $\alpha=0.70$ & 5 &18&  6.71 & 8.36 \\ \hline
DPD  & $\lambda=0.10$ & 9 &26& 10.57 & 9.62 \\ 
    & $\lambda=0.20$ & 9 &23&   9.65 & 9.51 \\ 
    & $\lambda=0.30$ & 9 &23&    8.99 & 9.43 \\ 
    & $\lambda=0.40$ & 9 &22&   8.30 & 9.32 \\ 
    & $\lambda=0.50$ & 9 &19&    7.87 & 9.16 \\ 
    & $\lambda=0.60$ & 7 &20&    7.52 & 8.99 \\ 
    & $\lambda=0.70$ & 7 &17&   7.26 & 8.96 \\ \hline
\end{tabular}
\end{table}

In Castilla et al. (2018), another family of divergence-based estimators was introduced as a robust alternative to the MLE in the PLRM. These estimators, named minimum DPD estimators, are also parametrized by a tuning parameter, let's say, $\lambda\geq 0$, and contain the MLE as a particular case for $\lambda=0$. For more details about minimum DPD estimators, see Ghosh and Basu (2013, 2015). A question that may arise here is why to use minimum RP estimators instead of minimum DPD estimators. In this regard, the performance of both families of estimators is compared in the following way: for each pair of estimators in the grid $\alpha \in [0,0.7]$, $\lambda\in [0,0.7]$, we compute the number of datasets where the classification performance of minimum RP estimators surpasses that of minimum DPD estimators, minus the number of datasets where the classification performance of minimum DPD estimators surpasses that of minimum RP estimators. Thus, if we have a positive (negative) value for a specific pair of estimators, it means that the minimum RP estimator outperforms (is outperformed by) the minimum DPD estimator. Results are presented in Figure \ref{fig:DIABETES_heat}. Note that the use of minimum RP estimators with a moderate to high value of the tuning parameter appears to be a better choice compared to using minimum DPD estimators for any value of the tuning parameter considered here. Further, some values relating the misclassification errors are added to Table \ref{table:DiabetesCOMP}, which highlight again the better performance of minimum RP estimators both for the unaltered and the contaminated datasets.  In any case, both families of estimators are shown to be necessary robust alternatives to MLE ($\alpha=\lambda=0$).

\section{Final remarks \label{sec:FR}}

In this study, we introduced a novel set of estimators, known as minimum RP estimators, which provide a robust alternative to the traditional MLE in the context of the PLRM. This family of estimators is parametrized by a tuning parameter $\alpha\geq0$, and includes the MLE as a special case when $\alpha=0$. We derived the estimating equations and asymptotic distribution for these estimators. Additionally, we proposed a family of Wald-type tests for linear hypothesis testing and developed the influence function of the proposed estimators and tests. An exhaustive simulation study and a practical example using real-world data were conducted to illustrate the robustness of these proposed statistics against misclassification errors. Notably, while the MLE is the most efficient estimator for non-contaminated data, the minimum RP estimators with $\alpha>0$ exhibit superior performance in the presence of such errors.

Expanding this approach to the PLRM with complex survey design could be particularly valuable for analyzing demographic and health surveys. This presents a significant problem worth exploring in the near future.  Finally, we hope that this study could potentially allow for more detailed examinations and robust estimations in diverse statistical contexts.\\

\noindent \textbf{Conflicts of interest:} The author declares that there are no conflicts of interest.


\appendix
\section{Proof of results \label{app:proofsWald}}
\subsection{Proof of Theorem \ref{th:WaldConsistent}}
Let us define
\begin{align*}
l_{\boldsymbol{\beta}^{\ast}}(\boldsymbol{\beta}^{\ast \ast})=\left(\boldsymbol{L}\boldsymbol{\beta}^{\ast}-\boldsymbol{l} \right)^T \left(\boldsymbol{L} \boldsymbol{V}_{\alpha}(\boldsymbol{\beta}^{\ast \ast})\boldsymbol{L}^T\right)^{-1}\left(\boldsymbol{L}\boldsymbol{\beta}^{\ast}-\boldsymbol{l} \right).
\end{align*}
Therefore, $nl_{\widehat{\boldsymbol{\beta}}_{\alpha}}(\widehat{\boldsymbol{\beta}}_{\alpha})=W_n(\widehat{\boldsymbol{\beta}}_{\alpha})$. As \ $\widehat{\boldsymbol{\beta}}_{\alpha} \xrightarrow[n \to \infty]{P}\boldsymbol{\beta}^1$, \ $l_{\widehat{\boldsymbol{\beta}}_{\alpha}}(\boldsymbol{\beta}^1)$ and $l_{\boldsymbol{\beta}^1}(\boldsymbol{\beta}^1)$ have the same asymptotic distribution. A first order Taylor expansion of $l_{\widehat{\boldsymbol{\beta}}_{\alpha}}(\boldsymbol{\beta})$ at $\widehat{\boldsymbol{\beta}}_{\alpha}$  around $\boldsymbol{\beta}^1$ gives
\begin{align*}
l_{\widehat{\boldsymbol{\beta}}_{\alpha}}(\widehat{\boldsymbol{\beta}}_{\alpha})=l_{\widehat{\boldsymbol{\beta}}_{\alpha}}(\boldsymbol{\beta}^1)+\left.\frac{\partial l_{\widehat{\boldsymbol{\beta}}_{\alpha}}(\boldsymbol{\beta})}{\partial \boldsymbol{\beta}^T} \right|_{\boldsymbol{\beta}=\boldsymbol{\beta}^1}(\widehat{\boldsymbol{\beta}}_{\alpha}-\boldsymbol{\beta}^1)+o_p(||\widehat{\boldsymbol{\beta}}_{\alpha}-\boldsymbol{\beta}^1 ||).
\end{align*}
Taking into account that $\sqrt{n} o_p(||\widehat{\boldsymbol{\beta}}_{\alpha}-\boldsymbol{\beta}^1 ||)=o_p(1)$, we have
\begin{align*}
\sqrt{n}\left( l_{\widehat{\boldsymbol{\beta}}_{\alpha}}(\widehat{\boldsymbol{\beta}}_{\alpha})-l_{\widehat{\boldsymbol{\beta}}_{\alpha}}(\boldsymbol{\beta}^1)\right)\xrightarrow[n\to \infty]{L} \mathcal{N}\left(0, \left.\frac{\partial l_{\widehat{\boldsymbol{\beta}}_{\alpha}}(\boldsymbol{\beta})}{\partial \boldsymbol{\beta}^T} \right|_{\boldsymbol{\beta}=\boldsymbol{\beta}^1}
\boldsymbol{V}_{\alpha}(\boldsymbol{\beta}^1)\left.\frac{\partial l_{\widehat{\boldsymbol{\beta}}_{\alpha}}(\boldsymbol{\beta})}{\partial \boldsymbol{\beta}} \right|_{\boldsymbol{\beta}=\boldsymbol{\beta}^1}\right).
\end{align*}
Let us denote
\begin{align}\label{eq:SigmainProof}
\sigma^2(\boldsymbol{\beta}^1)=\left.\frac{\partial l_{\widehat{\boldsymbol{\beta}}_{\alpha}}(\boldsymbol{\beta})}{\partial \boldsymbol{\beta}^T} \right|_{\boldsymbol{\beta}=\boldsymbol{\beta}^1}
\boldsymbol{V}_{\alpha}(\boldsymbol{\beta}^1)\left.\frac{\partial l_{\widehat{\boldsymbol{\beta}}_{\alpha}}(\boldsymbol{\beta})}{\partial \boldsymbol{\beta}} \right|_{\boldsymbol{\beta}=\boldsymbol{\beta}^1}.
\end{align}
Now, we have
\begin{align}\label{eq:PowerinProof}
P_{\boldsymbol{\beta}^1}(W_n(\widehat{\boldsymbol{\beta}}_{\alpha})>\chi^2_{r,\tau})&=P_{\boldsymbol{\beta}^1}\left(W_n(\widehat{\boldsymbol{\beta}}_{\alpha})-n >\chi^2_{r,\tau}-nl_{\boldsymbol{\beta}^1}(\boldsymbol{\beta}^1)\right)\\
&= P_{\boldsymbol{\beta}^1}\left(\frac{\sqrt{n}}{\sigma(\boldsymbol{\beta}^1)}\left(  l_{\widehat{\boldsymbol{\beta}}_{\alpha}}(\widehat{\boldsymbol{\beta}}_{\alpha})-l_{\widehat{\boldsymbol{\beta}}_{\alpha}}(\boldsymbol{\beta}^1) \right)>\frac{1}{\sigma(\boldsymbol{\beta}^1)}\left( \frac{\chi^2_{r,\tau}}{\sqrt{n}}-\sqrt{n}l_{\boldsymbol{\beta}^1}(\boldsymbol{\beta}^1)\right)  \right) \notag\\
& \approx 1-\phi \left( \frac{1}{\sigma(\boldsymbol{\beta}^1)}\left( \frac{\chi^2_{r,\tau}}{\sqrt{n}}-\sqrt{n}l_{\boldsymbol{\beta}^1}(\boldsymbol{\beta}^1)\right) \right), \notag
\end{align}
where $\phi(\cdot)$ is the distribution function of a standard normal distribution. Then, our result follows
\begin{align*}
\lim_{n \to \infty} P_{\boldsymbol{\beta}^1}(W_n(\widehat{\boldsymbol{\beta}}_{\alpha})>\chi^2_{r,\tau})=1.
\end{align*}

\bigskip

\subsection{Proof of Proposition \ref{th:PowerSize}}

From Equation (\ref{eq:PowerinProof}) we know that

\begin{align*}
\pi_{W_n(\widehat{\boldsymbol{\beta}}_{\alpha})}(\boldsymbol{\beta}^1) =P_{\boldsymbol{\beta}^1}(W_n(\widehat{\boldsymbol{\beta}}_{\alpha})>\chi^2_{r,\tau})& \approx 1-\phi \left( \frac{1}{\sigma(\boldsymbol{\beta}^1)}\left( \frac{\chi^2_{r,\tau}}{\sqrt{n}}-\sqrt{n}l_{\boldsymbol{\beta}^1}(\boldsymbol{\beta}^1)\right) \right),
\end{align*}
where $\sigma^2(\boldsymbol{\beta}^1)$ is given in (\ref{eq:SigmainProof}). Then, the result follows straightforward.

\bigskip

\section{Study of the boundedness of the influence functions in presence of outlying predictors \label{app:Brobust}}

 Here, we prove the non-B-robustness of minimum RP estimators following a similar approach to that of Miron et al. (2022). Note that prove that the influence function is unbounded reduces to prove that its score function 
 
 $$
\boldsymbol{\Psi}_{\alpha}(\boldsymbol{\beta})=\frac{1}{\left(\Gamma(\alpha,\boldsymbol{\beta})\right)^{\frac{\alpha}{1+\alpha}}}\boldsymbol{\Delta}^*(\boldsymbol{\pi}(\boldsymbol{\beta})) \text{diag}^{\alpha-1}\left(\boldsymbol{\pi}(\boldsymbol{\beta})\right)\left[\boldsymbol{y}-\frac{\Upsilon(\alpha,\boldsymbol{\beta})}{\Gamma(\alpha,\boldsymbol{\beta})}\boldsymbol{\pi}(\boldsymbol{\beta})\right]\otimes \boldsymbol{x}
$$
is unbounded. Miron et al. (2022) proved  that there exists a sequence $\boldsymbol{x}_n$ such that

$$
||\boldsymbol{x}_n||_2\to \infty \quad \text{and} \quad \pi_1(\boldsymbol{x}_n,\boldsymbol{\beta})=1/2.
$$
Now, we are going show that in such a case, when the response variable falls into the first category, the sequence $\boldsymbol{\Psi}_{\alpha}(\boldsymbol{\beta},\boldsymbol{x}_n,\boldsymbol{y}=\boldsymbol{e}_1)$ goes to infinity, meaning that $\boldsymbol{\Psi}_{\alpha}(\boldsymbol{\beta})$ is not bounded and therefore the influence function is unbounded too. We have that $\boldsymbol{\Psi}_{\alpha}(\boldsymbol{\beta})=\tilde{\boldsymbol{\Psi}}_{\alpha}(\boldsymbol{\beta}) \otimes \boldsymbol{x}$, with 

$$\tilde{\boldsymbol{\Psi}}_{\alpha}(\boldsymbol{\beta}) =\frac{1}{\left(\Gamma(\alpha,\boldsymbol{\beta})\right)^{\frac{\alpha}{1+\alpha}}}\boldsymbol{\Delta}^*(\boldsymbol{\pi}(\boldsymbol{\beta})) \text{diag}^{\alpha-1}\left(\boldsymbol{\pi}(\boldsymbol{\beta})\right)\left[\boldsymbol{y}-\frac{\Upsilon(\alpha,\boldsymbol{\beta})}{\Gamma(\alpha,\boldsymbol{\beta})}\boldsymbol{\pi}(\boldsymbol{\beta})\right],$$
and $||\boldsymbol{\Psi}_{\alpha}(\boldsymbol{\beta}) ||_2=||\tilde{\boldsymbol{\Psi}}_{\alpha}(\boldsymbol{\beta})||_2 \times ||\boldsymbol{x}||_2$. Now, after some computations, it can be shown that for the first coordinate of $\tilde{\boldsymbol{\Psi}}_{\alpha}(\boldsymbol{\beta},\boldsymbol{x}_n,\boldsymbol{y}=\boldsymbol{e}_1)$ is

\begin{align*}
\tilde{\boldsymbol{\Psi}}_{\alpha}(\boldsymbol{\beta},\boldsymbol{x}_n,\boldsymbol{y}=\boldsymbol{e}_1)_1&=\frac{\pi_1^{\alpha+1}(\boldsymbol{x}_n,\boldsymbol{\beta})}{\left(\Gamma(\alpha,\boldsymbol{\beta})\right)^{\frac{\alpha}{1+\alpha}}}\left[1-\frac{\Upsilon(\alpha,\boldsymbol{\beta})}{\Gamma(\alpha,\boldsymbol{\beta})}\left(\pi_1(\boldsymbol{x}_n,\boldsymbol{\beta})-\frac{1}{\pi_1^{\alpha}(\boldsymbol{x}_n,\boldsymbol{\beta})}\sum_{j=2}^{d+1}\pi_j^{\alpha+1}(\boldsymbol{x}_n,\boldsymbol{\beta}) \right) \right]\\
&=\frac{1}{2^{\alpha+1}\left(\Gamma(\alpha,\boldsymbol{\beta})\right)^{\frac{\alpha}{1+\alpha}}}\left[1-\frac{\Upsilon(\alpha,\boldsymbol{\beta})}{\Gamma(\alpha,\boldsymbol{\beta})}\left(\frac{1}{2}-2^{\alpha}\sum_{j=2}^{d+1}\pi_j^{\alpha+1}(\boldsymbol{x}_n,\boldsymbol{\beta}) \right) \right].
\end{align*}
But, it can be shown how, if $\pi_1(\boldsymbol{x}_n,\boldsymbol{\beta})=1/2$, 
$$\dfrac{\Upsilon(\alpha,\boldsymbol{\beta})}{\Gamma(\alpha,\boldsymbol{\beta})}\in [1,2) \quad \text{and} \quad \left(\frac{1}{2}-2^{\alpha}\sum_{j=2}^{d+1}\pi_j^{\alpha+1}(\boldsymbol{x}_n,\boldsymbol{\beta}) \right)\in \left[0,\frac{1}{2}\right),$$
therefore

$$
\tilde{\boldsymbol{\Psi}}_{\alpha}(\boldsymbol{\beta},\boldsymbol{x}_n,\boldsymbol{y}=\boldsymbol{e}_1)_1\geq K_{\alpha},
$$
for a positive constant $K_{\alpha}$. Thus
$$
||\boldsymbol{\Psi}_{\alpha}(\boldsymbol{\beta},\boldsymbol{x}_n,\boldsymbol{y}=\boldsymbol{e}_1) ||_2 \geq K_{\alpha} \times ||\boldsymbol{x}_n ||_2 \to \infty.
$$


\end{document}